\documentclass[12pt,preprint]{aastex}
\usepackage{graphicx}
\usepackage{lscape}

\newcommand{\pivec}{\mbox{\boldmath $\pi$}}
\newcommand{\muvec}{\mbox{\boldmath $\mu$}}
\newcommand{\nuvec}{\mbox{\boldmath $\nu$}}
\newcommand{\uvec}{\mbox{\boldmath $u$}}

\lefthead{JUNG ET AL.}
\righthead{ }

\begin{document}
\title{
%%%$Spitzer$ Parallax of OGLE-2018-BLG-0596: A Cold Uranus around an M-dwarf
$Spitzer$ Parallax of OGLE-2018-BLG-0596: A Low-mass-ratio Planet around an M-dwarf
}

\author{
Youn~Kil~Jung$^{1,35}$, 
Andrew~Gould$^{1,2,3,35,37}$,
Andrzej~Udalski$^{4,36}$,
Takahiro~Sumi$^{5,38}$,
Jennifer~C.~Yee$^{6,35,37}$,
Yossi~Shvartzvald$^{7,35,37,39}$,
Weicheng~Zang$^{8,40}$,
Cheongho~Han$^{9,35}$,\\
and \\
% KMTNet -----------------------------------------------------------------------------
Michael~D.~Albrow$^{10}$, Sun-Ju~Chung$^{1,11}$, Kyu-Ha~Hwang$^{1}$, 
Yoon-Hyun~Ryu$^{1}$, In-Gu~Shin$^{1}$, Wei~Zhu$^{12}$, Sang-Mok~Cha$^{1,13}$, 
Dong-Jin~Kim$^{1}$, Hyoun-Woo~Kim$^{1}$, Seung-Lee~Kim$^{1,11}$, Chung-Uk~Lee$^{1,11}$, 
Dong-Joo~Lee$^{1}$, Yongseok~Lee$^{1,13}$, Byeong-Gon~Park$^{1,11}$, Richard~W.~Pogge$^{2}$ \\
(The KMTNet Collaboration) \\
% OGLE -------------------------------------------------------------------------------
Przemek~Mr{\'o}z$^{4}$, Micha{\l}~K.~Szyma{\'n}ski$^{4}$, Jan~Skowron$^{4}$, Radek~Poleski$^{2}$,
Igor~Soszy{\'n}ski$^{4}$, Pawe{\l}~Pietrukowicz$^{4}$, Szymon~Koz{\l}owski$^{4}$, Krzystof~Ulaczyk$^{14}$, 
Krzysztof~A.~Rybicki$^{4}$, Patryk~Iwanek$^{4}$, Marcin~Wrona$^{4}$ \\
(The OGLE Collaboration) \\
% Spitzer -------------------------------------------------------------------------------
Charles~A.~Beichman$^{15,39}$, Geoffery~Bryden$^{16,39}$, Sebastiano~Calchi~Novati$^{7,39}$, 
Sean~Carey$^{17}$, B.~Scott~Gaudi$^{2,39}$, Calen~B.~Henderson$^{7,39}$ \\
(The $Spitzer$ Team) \\
% MOA --------------------------------------------------------------------------------
Fumio~Abe$^{18}$, Richard~Barry$^{19}$, David~P.~Bennett$^{19,20}$, Ian~A.~Bond$^{21}$, 
Aparna~Bhattacharya$^{19,20}$, Martin~Donachie$^{22}$, Akihiko~Fukui$^{23,24}$, Yuki~Hirao$^{5}$,
Yoshitaka~Itow$^{18}$, Iona~Kondo$^{5}$, Naoki~Koshimoto$^{25,26}$, Man~Cheung~Alex~Li$^{22}$, 
Yutaka~Matsubara$^{18}$, Shota~Miyazaki$^{5}$, Yasushi~Muraki$^{18}$, Masayuki~Nagakane$^{5}$, 
Cl{\'e}ment~Ranc$^{19}$, Nicholas~J.~Rattenbury$^{22}$, Haruno~Suematsu$^{5}$, Denis~J.~Sullivan$^{27}$,
Daisuke Suzuki$^{28}$, Paul~J.~Tristram$^{29}$, Atsunori~Yonehara$^{30}$ \\
(The MOA Collaboration) \\
% UKIRT -------------------------------------------------------------------------------
Savannah~Jacklin$^{31}$, Matthew~T.~Penny$^{2,40}$, Keivan~G.~Stassun$^{31}$ \\
(The UKIRT Microlensing Team) \\
% CFHT -------------------------------------------------------------------------------
Pascal~Fouqu\'e$^{32,33}$, Shude~Mao$^{8,34}$, and Tianshu~Wang$^{8}$ \\
(The CFHT Microlensing Collaboration)
}

\bigskip\bigskip
\affil{$^{1}$Korea Astronomy and Space Science Institute, Daejon 34055, Republic of Korea}
\affil{$^{2}$Department of Astronomy, Ohio State University, 140 W. 18th Ave., Columbus, OH 43210, USA}
\affil{$^{3}$Max-Planck-Institute for Astronomy, K$\rm \ddot{o}$nigstuhl 17, 69117 Heidelberg, Germany}
\affil{$^{4}$Warsaw University Observatory, Al. Ujazdowskie 4, 00-478 Warszawa, Poland}
\affil{$^{5}$Department of Earth and Space Science, Graduate School of Science, Osaka University, Toyonaka, Osaka 560-0043, Japan}
\affil{$^{6}$Center for Astrophysics $|$ Harvard \& Smithsonian, 60 Garden St., Cambridge, MA 02138, USA}
\affil{$^{7}$IPAC, Mail Code 100-22, Caltech, 1200 E. California Blvd., Pasadena, CA 91125, USA}
\affil{$^{8}$Physics Department and Tsinghua Centre for Astrophysics, Tsinghua University, Beijing 100084, China}
\affil{$^{9}$Department of Physics, Chungbuk National University, Cheongju 28644, Republic of Korea}
\affil{$^{10}$University of Canterbury, Department of Physics and Astronomy, Private Bag 4800, Christchurch 8020, New Zealand}
\affil{$^{11}$University of Science and Technology, Korea, 217 Gajeong-ro Yuseong-gu, Daejeon 34113, Korea}
\affil{$^{12}$Canadian Institute for Theoretical Astrophysics, University of Toronto, 60 St George Street, Toronto, ON M5S 3H8, Canada}
\affil{$^{13}$School of Space Research, Kyung Hee University, Yongin 17104, Republic of Korea}
\affil{$^{14}$Department of Physics, University of Warwick, Gibbet Hill Road, Coventry, CV4 7AL, UK}
\affil{$^{15}$NASA Exoplanet Science Institute, MS 100-22, California Institute of Technology, Pasadena, CA 91125, USA}
\affil{$^{16}$Jet Propulsion Laboratory, California Institute of Technology, 4800, Oak Grove Drive, Pasadena, CA 91109, USA}
\affil{$^{17}$Spitzer Science Center, MS 220-6, California Institute of Technology, Pasadena, CA, USA}
\affil{$^{18}$Institute for Space-Earth Environmental Research, Nagoya University, Nagoya 464-8601, Japan}
\affil{$^{19}$Code 667, NASA Goddard Space Flight Center, Greenbelt, MD 20771, USA}
\affil{$^{20}$Department of Astronomy, University of Maryland, College Park, MD 20742, USA}
\affil{$^{21}$Institute of Natural and Mathematical Science, Massey University, Auckland 0745, New Zealand}
\affil{$^{22}$Department of Physics, University of Auckland, Private Bag 92019, Auckland, New Zealand}
\affil{$^{23}$Department of Earth and Planetary Science, Graduate School of Science, The University of Tokyo, 7-3-1 Hongo, Bunkyo-ku, Tokyo 113-0033, Japan}
\affil{$^{24}$Instituto de Astrof{\'i}sica de Canarias, V{\'i}a L{\'a}ctea s/n, E-38205 La Laguna, Tenerife, Spain}
\affil{$^{25}$Department of Astronomy, Graduate School of Science, The University of Tokyo, 7-3-1 Hongo, Bunkyo-ku, Tokyo 113-0033, Japan}
\affil{$^{26}$National Astronomical Observatory of Japan, 2-21-1 Osawa, Mitaka, Tokyo 181-8588, Japan}
\affil{$^{27}$School of Chemical and Physical Science, Victoria University, Wellington, New Zealand}
\affil{$^{28}$Institute of Space and Astronautical Science, Japan Aerospace Exploration Agency, Kanagawa 252-5210, Japan}
\affil{$^{29}$University of Canterbury Mt. John Observatory, P.O. Box 56, Lake Tekapo 8770, New Zealand}
\affil{$^{30}$Department of Physics, Faculty of Science, Kyoto Sangyo University, Kyoto 603-8555, Japan}
\affil{$^{31}$Vanderbilt University, Department of Physics \& Astronomy, PMB 401807, 2301 Vanderbilt Place, Nashville, TN 37235, USA}
\affil{$^{32}$CFHT Corporation, 65-1238 Mamalahoa Hwy, Kamuela, Hawaii 96743, USA}
\affil{$^{33}$Universit\'e de Toulouse, UPS-OMP, IRAP, Toulouse, France}
\affil{$^{34}$National Astronomical Observatories, Chinese Academy of Sciences, A20 Datun Rd., Chaoyang District, Beijing 100012, China}
% ------------------------------------------------------------------------------------------------------
\footnotetext[35]{The KMTNet Collaboration.}
\footnotetext[36]{The OGLE Collaboration.}
\footnotetext[37]{The $Spitzer$ Team.}
\footnotetext[38]{The MOA Collaboration.}
\footnotetext[39]{The UKIRT Microlensing Team.}
\footnotetext[40]{The CFHT Mircolensing Collaboration.}

\begin{abstract}
We report the discovery of a $Spitzer$ microlensing planet OGLE-2018-BLG-0596Lb, 
with preferred planet-host mass ratio $q \sim 2\times10^{-4}$. 
The planetary signal, which is characterized by a short $(\sim 1~{\rm day})$ ``bump'' 
on the rising side of the lensing light curve, was densely covered by ground-based 
surveys. We find that the signal can be explained by a bright source that fully envelops 
the planetary caustic, i.e., a ``Hollywood'' geometry. Combined with the source proper 
motion measured from $Gaia$, the $Spitzer$ satellite parallax measurement makes it 
possible to precisely constrain the lens physical parameters. The preferred solution, 
in which the planet perturbs the minor image due to lensing by the host, yields a 
Uranus-mass planet with a mass of $M_{\rm p} = 13.9\pm1.6~M_{\oplus}$ orbiting a mid 
M-dwarf with a mass of $M_{\rm h} = 0.23\pm0.03~M_{\odot}$. There is also a second 
possible solution that is substantially disfavored but cannot be ruled out, for which 
the planet perturbs the major image. The latter solution yields 
$M_{\rm p} = 1.2\pm0.2~M_{\oplus}$ and $M_{\rm h} = 0.15\pm0.02~M_{\odot}$. 
By combining the microlensing and $Gaia$ data together with a Galactic model, 
we find in either case that the lens lies on the near side of the Galactic bulge 
at a distance $D_{\rm L} \sim 6\pm1~{\rm kpc}$. Future adaptive optics observations 
may decisively resolve the major image/minor image degeneracy.
\end{abstract}
%%\keywords{binaries: general -- gravitational lensing: micro -- planetary systems}

%%The event was densely 
%%covered by ground-based surveys, in which the planetary perturbation takes the form 
%%of a weak, one-day ``bump'' on the rising side of the light curve. We find that 
%%the perturbation can be explained by a bright source that fully envelops the planetary 

\section{Introduction}

In microlensing events, the principal observable connected to the physical properties 
of the lens is the Einstein timescale $t_{\rm E}$. However, the timescale results from 
a combination of the lens mass $M$ and the lens-source relative proper motion 
$\mu_{\rm rel}$ and parallax $\pi_{\rm rel}$, i.e.,  
\begin{equation}
t_{\rm E} = {\theta_{\rm E} \over \mu_{\rm rel}};~~~~~\theta_{\rm E} = \sqrt{{\kappa}M{\pi_{\rm rel}}},  
\label{eq1}
\end{equation}
where $\theta_{\rm E}$ is the angular Einstein radius and 
\begin{equation}
\kappa = {4{\rm G} \over {c^2{\rm au}}} \sim 8.14~{\rm mas}\,M_{\odot}^{-1};~~~~~
\pi_{\rm rel} = {\rm au}\left({{1 \over D_{\rm L}} - {1 \over D_{\rm S}}}\right).
\label{eq2}
\end{equation}
Here, $D_{\rm L}$ and $D_{\rm S}$ are the lens and the source distances, respectively. 
Therefore, it is difficult to uniquely constrain the lens physical parameters from 
the timescale alone. To resolve this $(M, \mu_{\rm rel}, \pi_{\rm rel})$ degeneracy 
requires measuring two additional quantities: $\theta_{\rm E}$ and the microlens 
parallax $\pivec_{\rm E}$. The measurements of these two quantities enable one to 
determine the physical parameters through the relations \citep{gould00} 
\begin{equation}
M = {\theta_{\rm E} \over {\kappa \pi_{\rm E}}};~~~~~
\pi_{\rm rel} = {\theta_{\rm E}\pi_{\rm E}};~~~~~
\muvec_{\rm rel} = {\theta_{\rm E} \over t_{\rm E}}{\pivec_{\rm E} \over \pi_{\rm E}}. 
\label{eq3}
\end{equation}
Additionally, if the source proper motion $\muvec_{\rm S}$ and parallax 
$\pi_{\rm S} = {\rm au}/D_{\rm S}$ are independently estimated, the $\theta_{\rm E}$ and 
$\pivec_{\rm E}$ measurements allow one to infer the phase space coordinates of the lens 
system by    
\begin{equation}
\muvec_{\rm L} = \muvec_{\rm rel} + \muvec_{\rm S};~~~~~
\pi_{\rm L} = \pi_{\rm rel} + \pi_{\rm S},
\label{eq4}
\end{equation}
where $\muvec_{\rm L}$ and $\pi_{\rm L} = {\rm au}/D_{\rm L}$ are the lens proper motion 
and parallax, respectively.

As summarized by \citet{zhu15}, there are several approaches for the measurement of 
$\theta_{\rm E}$, but the most common is to detect the deviation in the observed light 
curve induced by the extended nature of source stars, i.e., finite-source effects. Such 
a deviation arises when the source is placed in or near the region where the lensing 
magnification of a point-like source would diverge to infinity (i.e., a caustic). 
The detection of the finite-source effect usually returns the source radius normalized 
to the Einstein radius, $\rho = \theta_{*}/\theta_{\rm E}$, where $\theta_{*}$ is the 
angular radius of the source. Because $\theta_{*}$ is routinely measured from the 
additional information of the source color and magnitude \citep{yoo04}, one can 
determine $\theta_{\rm E}$ provided that $\rho$ is measured from the light curve.

The microlens parallax can be measured through the annual microlens parallax effect. 
This effect arises from the orbital acceleration of Earth, which displaces the position 
of an observer relative to rectilinear lens-source motion \citep{gould92}. However, 
the measurement of $\pivec_{\rm E}$ in this single accelerating frame is usually difficult 
because the change of the observer's position during typical microlensing 
events $(t_{\rm E} < {\rm yr}/2\pi)$ is quite minor. As a result, the sample of events 
with $\pivec_{\rm E}$ measured from the annual parallax effect is relatively small, and 
they are biased toward events with long timescales (e.g., \citealt{jung19b}) and/or 
events produced by nearby lenses (e.g., \citealt{jung18a}).

The alternative way to measure $\pivec_{\rm E}$ is to use a satellite in a heliocentric orbit: 
the space-based microlens parallax effect. For typical lensing events, the displacement of 
the satellite from Earth comprises a substantial fraction of the projected Einstein radius 
$\tilde{r}_{\rm E} = {\rm au}/\pi_{\rm E} \sim 10~{\rm au}$. In this case, the lensing 
light curves simultaneously observed from Earth and the satellite can appear to be different 
because the time-dependent lens-source separation seen from the two observers can be different. 
Then, one can measure the microlens parallax by comparing these two light curves. This idea was 
first proposed by \citet{refsdal66} a half century ago, and the first such $\pivec_{\rm E}$ 
measurement was made by \citet{dong07}, in which they analyzed the event OGLE-2005-SMC-001 
by using both ground-based and $Spitzer$ observations. Subsequently, about a thousand 
microlensing events have been observed through the $Spitzer$ microlensing campaign 
\citep{prop2013,prop2014,prop2015a,prop2015b,prop2016,prop2018} 
in order to measure their microlens parallaxes. Combined with ground-based observations, 
these $\pivec_{\rm E}$ measurements have provided a unique opportunity to probe a variety of 
astrophysical populations, including binary brown dwarfs \citep{albrow18}, 
single-mass objects \citep{zhu16,chung17,shin18,shvartzvald19}, and 
planetary systems \citep{udalski15b,street16,shvartzvald17b,ryu18,calchi18,calchi19}.

Here, we analyze the microlensing event OGLE-2018-BLG-0596 and 
report the discovery of a low-mass-ratio planet orbiting a mid M-dwarf, i.e., 
with a preferred mass ratio of $q \sim 2\times10^{-4}$. The event was observed by 
several ground-based surveys and $Spitzer$, and the proper motion of the 
microlensed source was independently measured from $Gaia$. The ground-based 
observations clearly show a short-term anomaly in the rising part of the light curve, 
from which the presence of the planet is inferred. Moreover, the parallax 
measurement from $Spitzer$ and the proper motion from $Gaia$ allow us to precisely 
constrain the lens physical properties.

\section{Observation}

\subsection{Ground-based Observations}

OGLE-2018-BLG-0596 is at $({\rm RA}, {\rm Dec})_{\rm J2000} = $(17:56:13.33, $-29$:11:56.7), 
corresponding to $(l,b) = (0.96, -2.13)$ in Galactic coordinates. It was discovered as a 
probable microlensing event by the Optical Gravitational Lensing Experiment \citep[OGLE:][]{udalski15a} 
survey, and announced on 2018 April 15 through its Early Warning System \citep{udalski03}. 
The event is in the OGLE-IV field BLG505, for which OGLE observations were conducted with 
a one hour cadence using the 1.3 m Warsaw telescope located at Las Campanas in Chile. 

The Microlensing Observations in Astrophysics \citep[MOA:][]{sumi03} survey independently 
discovered this event on May 15 and named it as MOA-2018-BLG-145. The MOA observations were 
taken using the 1.8 m MOA-II telescope located at Mt. John Observatory in New Zealand. 
The MOA observation cadence for the event is 15 minutes.

The event was also independently discovered by the Korea Microlensing Telescope Network \citep[KMTNet:][]{kim16} 
by employing their post-season event finder algorithm \citep{kim18}, and it was cataloged as KMT-2018-BLG-0945. 
The KMTNet survey used three 1.6 m telescopes positioned at the Cerro Tololo Interamerican Observatory, 
Chile (KMTC), the South African Astronomical Observatory, South Africa (KMTS), and the Siding Spring 
Observatory, Australia (KMTA). The KMTNet observations were conducted with a 30-minute cadence.  

The great majority of images were obtained in the $I$ band for OGLE and KMTNet and a customized 
$R$ band for MOA\footnote{MOA survey uses a custom wide-band filter referred as MOA-Red, 
corresponding to the combination of a standard $R$- and $I$-band.}, with some $V$-band images 
for the source color measurement. All of the survey data were reduced using the image subtraction 
methodology \citep{alard98}, specifically \citet{wozniak00} for OGLE, \citet{bond01} for MOA, 
and \citet{albrow09} for KMTNet.

In addition to the observations from these high-cadence surveys, the event was observed by 
two lower-cadence surveys. These surveys used, respectively, the 3.8 m United 
Kingdom Infrared Telescope \citep[UKIRT:][]{shvartzvald17a} and the 3.6 m Canada-France-Hawaii Telescope 
\citep[CFHT:][]{zang18} that are both located at the Mauna Kea Observatory in Hawaii. The UKIRT and CFHT 
observations for the event were carried out in the $H$ and $i$ band, respectively.
%%%%Reductions of UKIRT and CFHT data 
%%%%were completed using the methods of Shvartzvald et al. (2018) and Zang et al. (2018), respectively.  

\subsection{$Spitzer$ Observations}

On May 10, the KMTNet group noticed in KMTNet data reduced on the basis of the OGLE alert
that the event had shown an anomaly at ${\rm HJD}'(={\rm HJD} - 2450000) \sim 8243.5$. 
Because this anomaly occurred when the event was just $\sim 0.3$ mag brighter than its 
baseline, it was impossible to precisely determine the lensing parameters at that time. 
Nevertheless, they found from real-time modeling that the anomaly was likely to have been 
produced by a very low-mass companion to the primary lens, i.e., a planetary companion. 
In response to this potential importance, the $Spitzer$ team announced OGLE-2018-BLG-0596 
as a $Spitzer$ target on May 24. 
The $Spitzer$ observations for the event were initiated on July 4 (when it first became 
observable due to Sun-angle restrictions) with a cadence of 1 day. In total, 
36 images were taken during $8304 < {\rm HJD}' < 8341$. The data were reduced 
based on the methods presented by \citet{calchi15b}. 

%%%as a $Spitzer$ target on May 24 for the purpose of measuring its microlens parallax. 

\subsubsection{Is the Event Part of the $Spitzer$ Parallax Sample?}

The main goal of the $Spitzer$ microlensing campaign is to derive the
Galactic distribution of planets \citep{calchi15a,zhu17}. In order to 
have an unbiased sample, which is essential to achieve this goal, the events 
that are included in the experiment must follow strict selection protocols 
specified by \citet{yee15a}. Because OGLE-2018-BLG-0596 was selected as a 
$Spitzer$ target significantly after the planetary anomaly, naively it seems 
that it should immediately be excluded from the sample. However, \citet{yee15a} 
anticipated exactly this situation (an early planetary anomaly) and specified 
strict selection criteria under which these planets can be included in the sample 
while keeping it unbiased. For example, OGLE-2016-BLG-1190 \citep{ryu18} also had 
an early planetary anomaly and is part of the $Spitzer$ sample thanks to these
protocols.

\citet{yee15a} specified two classes of ``objective'' selection criteria under which 
an event might be included in the sample: rising events and events that already peaked, 
i.e., falling events. All events that pass these strict criteria {\it must} be observed 
by $Spitzer$. The time threshold between the two classes is $t_0=t_{\rm dec} - 2$ days, 
where $t_{0}$ is the time of maximum magnification and  $t_{\rm dec}$ is the time when 
$Spitzer$ observations are finalized before each observing week. In the case of OGLE-2018-BLG-0596, 
the first decision date was 2018 June 25, Monday UT 13:25, i.e., 
$t_{\rm dec}=8295.06$. Because the event already peaked more than two weeks earlier it should be 
considered under the criteria for falling events (Section 6.1 of \citealt{yee15a}).

The falling event criteria include six criteria (A1-A6). The first is simply 
the definition of a falling event, A1: $t_0>t_{\rm dec} - 2$ days, which in 
marginal cases needs to be carefully modeled but in the case of OGLE-2018-BLG-0596 
was clearly already satisfied. The second criterion is for the event to be in a 
relatively high cadence OGLE or KMT field, which as described in Section 2.1 is 
clearly satisfied. The third criterion requires that the event peaked brighter 
than A3: $I_{\rm peak} < 17~{\rm mag}$, which again is clearly satisfied.

The next three are model-dependent criteria and must be examined by (1) using 
the data that were available to the team at $t_{\rm dec}$ and (2) removing the 
signature of the planetary anomaly (i.e., excluding the data during $8240 < {\rm HJD'} < 8246$). 
In addition, these criteria require the evaluation of the magnification of a 
single-lens single-source (1L1S) model at two specific times, $t_{\rm next}$ 
and $t_{\rm fin}$ which are the time of the next (i.e., first) and last possible 
$Spitzer$ observations, respectively. We fit the event to a single lens event with 
the on-line OGLE, MOA, and KMT data that were available to the team by $t_{\rm dec}$, 
after excluding the anomalous region, and then checked the next criterion. 
We note that the $Spitzer$ team did this examination also immediately after 
$t_{\rm dec}$, and reached the same conclusions that we find below.

Criterion A4 requires that there will be significant magnification change during 
the observable $Spitzer$ window, A4: $A(t_{\rm next})-A(t_{\rm fin})>0.3$. We find 
$A(t_{\rm next})-A(t_{\rm fin})=0.29$. We note for completeness that the event easily 
passes Criterion A5 (that the event will be bright enough to observe from $Spitzer$) 
and Criterion A6 (that the event will undergo a significant change in magnitude during
the $Spitzer$ observations). However, because it fails A4, we conclude that OGLE-2018-BLG-0596 
{\it does not} meet the ``objective'' selection criteria and {\it cannot} be included 
in the $Spitzer$ sample.

\section{Analysis}

\begin{figure}[th]
%%\epsscale{0.9}
\epsscale{0.9}
\plotone{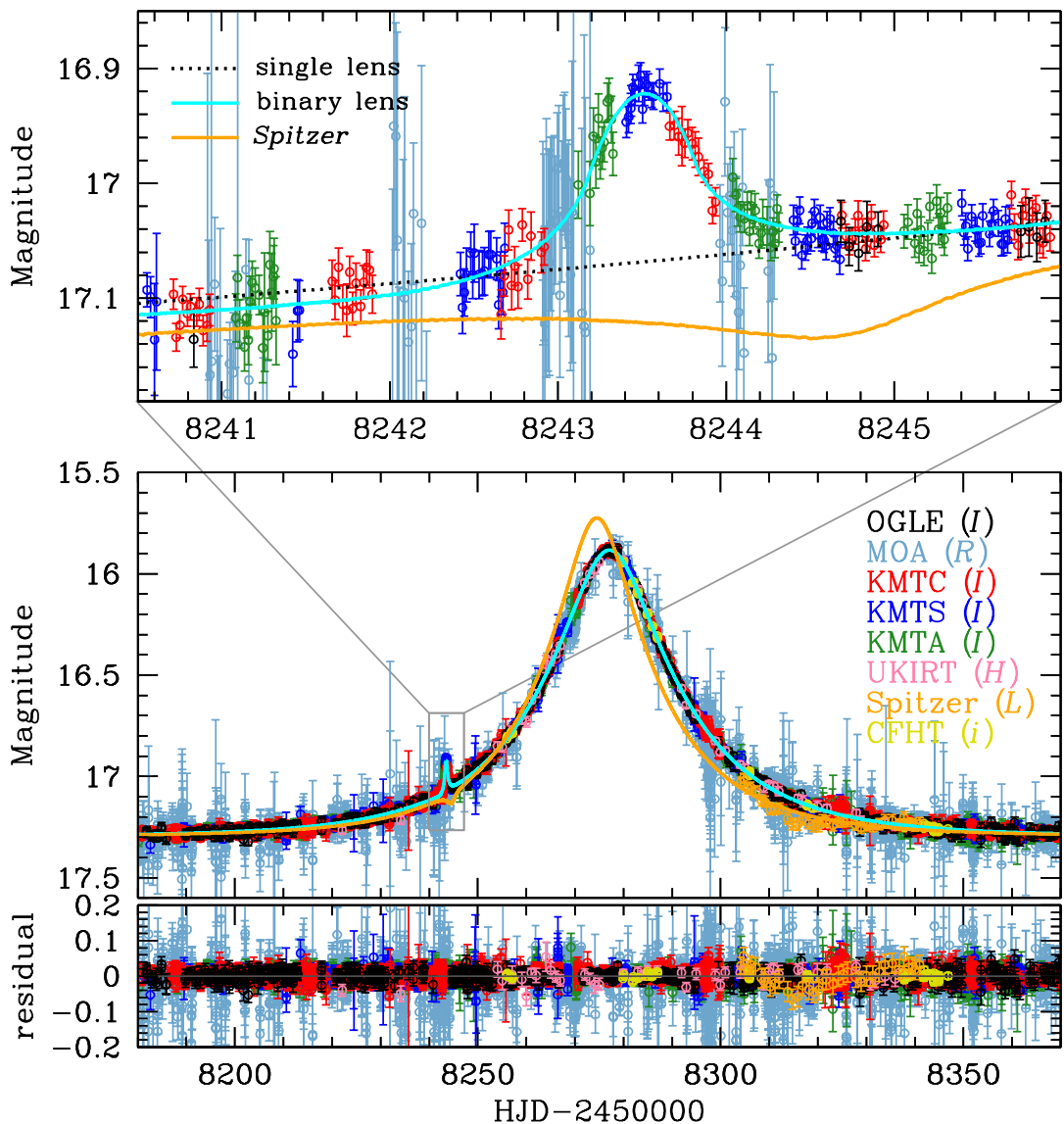}
\caption{\label{fig:one}
Light curve of OGLE-2018-BLG-0596. The upper panel shows 
the enlarged view of the anomaly centered at ${\rm HJD}' \sim 8243.5$.
The cyan and orange curves are the best-fit model curves for the ground- 
and space-based observations, respectively. The black dotted curve is 
the model curve obtained from the 1L1S interpretation.   
}
\end{figure}

Figure~\ref{fig:one} displays the light curve of OGLE-2018-BLG-0596 with the best-fit model. 
Ignoring the $Spitzer$ data, it primarily takes the symmetric form of a standard 
\citet{paczynski86} curve with a magnification $A_{\rm max} \sim 3.6$ at the peak. 
However, there is a short-lived, weak ``bump'' on the rising part of the light curve 
at ${\rm HJD}' \sim 8243.5$. This appearance could be produced by a ``Hollywood'' 
geometry \citep{gould97}, i.e., a small caustic that is substantially 
(or fully) enveloped by the source (e.g., \citealt{beaulieu06,hwang18}). 
Therefore, we begin by applying the binary-lens single-source (2L1S) 
interpretation to the event to explain the observed brightness variation.

\subsection{Ground-based Model}

We first model the light curve based on the data acquired from the ground-based 
observations. For our standard binary-lens model, we introduce seven non-linear parameters. 
(see the Appendix of \citealt{jung15} for graphical presentation of the parameters.) 
This includes three single-lens parameters $(t_{0}, u_{0}, t_{\rm E})$, 
three parameters for the binary companion $(s, q, \alpha)$, and one parameter for the source 
radius $\rho$. Here, $u_{0}$ is the impact parameter 
(in units of $\theta_{\rm E}$), $s$ is the companion-host projected separation 
(in units of $\theta_{\rm E}$), $q = M_{2}/M_{1}$ is their mass ratio, and 
$\alpha$ is their orientation angle with respect to $\muvec_{\rm rel}$. 
In addition, we introduce two flux parameters $(f_{\rm S}, f_{\rm B})_{i}$ 
for each data set in order to model the observed flux $f_{i}(t)$ as
\begin{equation}
f_{i}(t) = f_{{\rm S},i}A(t) + f_{{\rm B}, i},
\label{eq5}
\end{equation}
where $A(t)$ is the magnification given by the model and the subscripts 
``S'' and ``B'' denote the source and any blended light, respectively.

With these fitting parameters, we carry out a systematic analysis by following 
the procedure of \citet{jung15}. First, we estimate initial values of 
$(t_{0}, u_{0}, t_{\rm E}) = (8277.17, 0.28, 28.83~{\rm days})$ 
by fitting a 1L1S curve to the data set with the anomaly excluded. 
We also adopt an initial value of $\rho = 0.01$ based on $t_{\rm E}$ and 
the source brightness estimated from the fit. We next perform a grid 
search over $(s, q)$, in which $(s, q)$ are held fixed, 
while $(t_{0}, u_{0}, t_{\rm E}, \alpha, \rho)$ are sought based on a downhill approach. 
For this approach, as well as for determining the uncertainties of the parameters, 
we employ a Markov Chain Monte Carlo (MCMC) algorithm. For each set of fitting parameters, 
the lensing magnification is evaluated by inverse ray-shooting \citep{kayser86,schneider87} 
in the anomaly region and by semi-analytic approximations \citep{pejcha09,gould08} elsewhere. 
This model magnification is then used to fit the flux parameters $(f_{\rm S}, f_{\rm B})_{i}$ 
that minimize the $\chi^{2}$ of the observed flux $f_{i}(t)$.

\begin{figure}[th]
%%\epsscale{0.9}
\epsscale{0.9}
\plotone{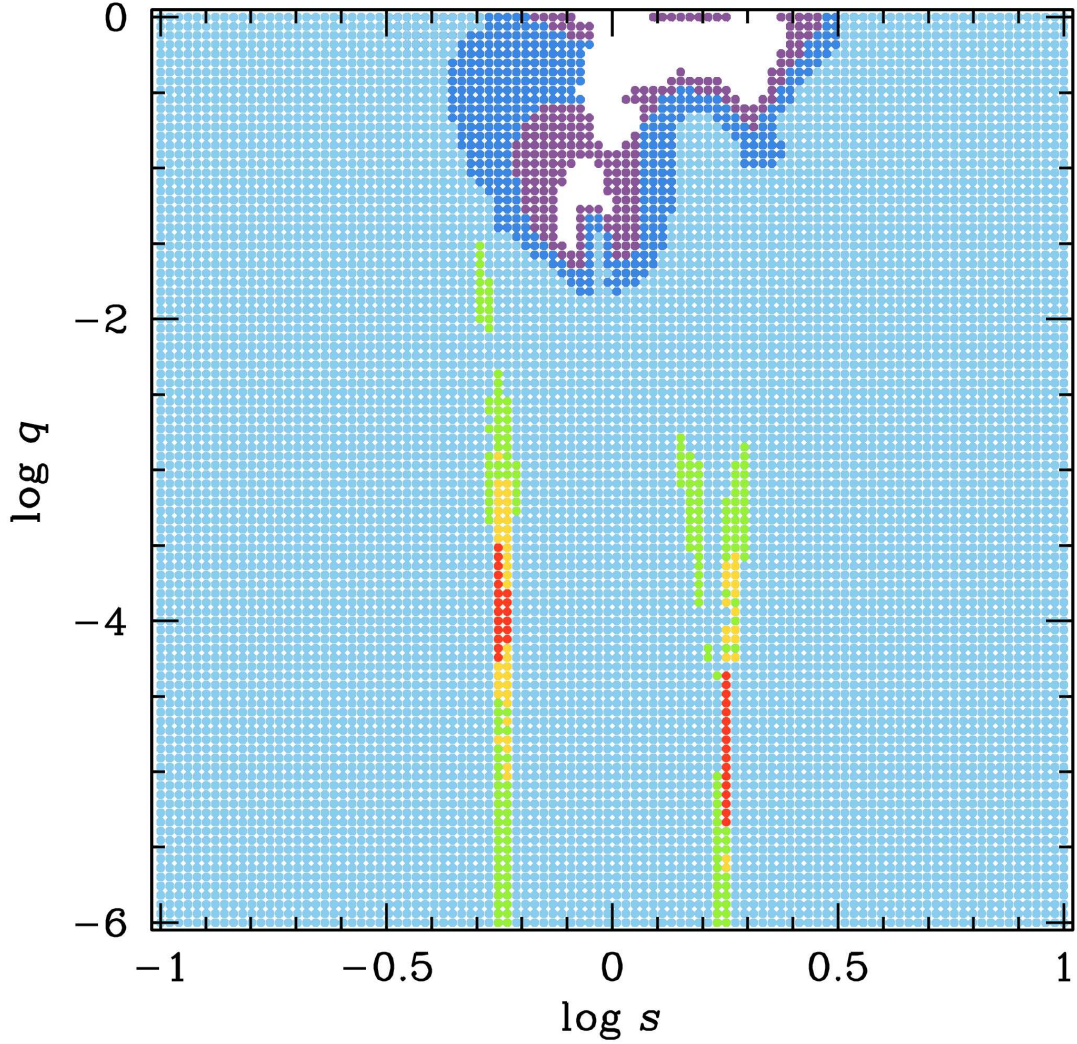}
\caption{\label{fig:two}
$\Delta\chi^{2}$ map in $({\rm log}~s, {\rm log}~q)$ space drawn from the grid search.
The space is equally divided on a $(100\times100)$ grid with ranges of $-1 < {\rm log}\,s < 1$ and 
$-6 < {\rm log}\,q < 0$, respectively. The contour is color coded by 
(red, yellow, green, light blue, blue, purple) for 
$\Delta\chi^2 < [(1\,n)^{2}, (2\,n)^{2}, (3\,n)^{2}, (4\,n)^{2}, (5\,n)^{2}, (6\,n)^{2}]$, where $n = 20$. 
}
\end{figure}

Figure~\ref{fig:two} displays the $\Delta\chi^{2}$ distribution on the $({\rm log}\,s, {\rm log}\,q)$ 
space acquired from the grid search. We identify two local minima, one with $s < 1$ (``Close'') 
and the other with $s > 1$ (``Wide''). We find that in both solutions, the lens system responsible for 
the weak ``bump'' is composed of two masses with $q \lesssim 10^{-4}$, implying that 
the lower-mass component is a planet. We then seed the local solutions 
into MCMCs and allow all fitting parameters to vary. The two standard solutions 
derived from this refinement process are given in Table~\ref{table:one} and~\ref{table:two}.

\subsection{Microlens Parallax and Lens Orbital Motion}

We now take into account the microlens parallax in order to simultaneously model the data 
obtained from the ground and $Spitzer$. This introduces two additional parameters 
$\pivec_{\rm E} = (\pi_{{\rm E},N}, \pi_{{\rm E},E})$, which represent the vector microlens 
parallax \citep{gould92}, i.e., 
\begin{equation}
\pivec_{\rm E} = {\pi_{\rm rel} \over \theta_{\rm E}}{\muvec_{\rm rel} \over \mu_{\rm rel}}.
\label{eq6}
\end{equation}
Then, the parallax parameters are approximately related to the offset $\Delta\uvec = (\Delta\beta, \Delta\tau)$ 
between the two light curves observed from the ground and $Spitzer$, i.e.,  
\begin{equation}
\pivec_{\rm E} = {{\rm au} \over D_{\perp}}(\Delta\beta, \Delta\tau);~~~~~
\Delta\beta = u_{0,\oplus} - u_{0,{\rm sat}};~~~~~
\Delta\tau = {{t_{0,\oplus} - t_{0,{\rm sat}}} \over t_{\rm E}},
\label{eq7}
\end{equation}
where $D_{\perp}$ is the projected Earth-$Spitzer$ separation and 
$(\Delta\beta, \Delta\tau)$ represent the components of the lens-source
separation vector that are perpendicular to and parallel with the source
trajectory, respectively. For single-lens events, the perpendicular 
offset $\Delta\beta$ generally suffers from a fourfold degeneracy, 
\begin{equation}
\Delta\beta = {\pm}u_{0,\oplus} - {\pm}u_{0,{\rm sat}},
\label{eq8}
\end{equation}
due to the rotational symmetry of the lensing magnification about the lens \citep{refsdal66,gould94}. 
These four possible solutions are usually denoted by $(+, +)$, $(-, -)$, $(+, -)$, and $(-, +)$ 
depending on the signs of $u_{0,\oplus}$ and $u_{0,{\rm sat}}$. For binary lenses, however, 
the fourfold degeneracy persists only if the source trajectory is almost parallel to the binary 
axis, i.e., $\alpha \sim 0$ \citep{zhu15}, and otherwise is reduced to a twofold degeneracy: $(+,+)$ and $(-,-)$. 
We therefore expect that OGLE-2018-BLG-0596 may only suffer from a twofold degeneracy, 
but we need a detailed analysis to draw a definitive conclusion. 
%%$(\Delta\beta, \Delta\tau)$ are the perpendicular and parallel offset 
%%of the lens-source separations. For single-lens events, the perpendicular 

To conduct a systematic analysis, we first consider additional information extracted from 
the ground- and space-based observations. As shown in Figure~\ref{fig:one}, the $Spitzer$ data only 
cover the falling side of the event and do not cover the anomaly. In such a case, 
it is difficult to precisely constrain $\pivec_{\rm E}$ from the data alone. We therefore 
apply a color constraint on the $Spitzer$ source flux to improve the parallax measurement 
\citep{yee15b}. For this, we derive an $IHL$ color-color relation using the OGLE, UKIRT, and 
$Spitzer$ data based on the method described by \citet{calchi15b}\footnote{We note that 
the $Spitzer$ bandpass is centered at 3.6 $\mu$m, which we designate as the $L$ band.}. 
From a model-independent regression of $I$- and $H$-band data \citep{gould10}, we first measure the instrumental 
$(I-H)$ color of the source as $(I-H)_{\rm S} = 2.81 \pm 0.01$. We next construct 
$(I-H, I)$ and $(I-L, I)$ instrumental color-magnitude diagrams (CMD) by cross-matching 
field stars within $120''$ of the source. We then conduct the color-color regression 
on red giant stars $(15 < I < 18;~2.4 < (I-H) < 3.0)$ to confine the sample to the bulge 
population. From this, we find $(I - L) = 1.43(I-H) - 8.72$. We thereby derive 
$(I-L)_{\rm S} = -4.70\pm0.02$, where the instrumental $Spitzer$ magnitude is given by 
$L = 25 - 2.5{\rm log}\,f_{{\rm S}, Spitzer}$. We impose this color-constraint on 
the model by adding a $\chi^{2}$ penalty, i.e.,  
\begin{equation}
\chi^{2}_{{\rm penalty}} = {[2.5{\rm log}\,(R_{\rm model}/R_{\rm constraint})]^{2} \over \sigma^{2}_{\rm constraint}},
\label{eq9}
\end{equation}
where $R$ is the flux ratio between $I$ and $L$ band and $\sigma_{\rm constraint}$ 
is the uncertainty of $(I-L)_{\rm S}$.

Space-based observations can provide an opportunity not only to measure the microlens parallax 
but also to constrain the orbital motion of the binary lens \citep{han16}. As discussed by \citet{batista11}, 
the annual microlens parallax (due to Earth's orbital motion) can be partially degenerate with the lens orbital motion, 
and so the microlens parallax measured from a single accelerating frame can absorb the lens orbital motion. By contrast, 
the space-based microlens parallax does not suffer from this degeneracy because it is determined 
from the feature of the light curves from two different observatories. This enables 
one to break such degeneracy and detect the lens orbital motion in the ground-based light curve. 
Therefore, we also take into account the lens orbital effect. To account for this effect, 
we introduce two linearized parameters $(ds/dt, d\alpha/dt)$, which represent the relative 
velocity of the lens components projected onto the plane of the sky.

%%As discussed by \citet{han16}, space-based observations can provide an opportunity 
%%not only to measure the microlens parallax but also to constrain the orbital motion 
%%of the binary lens. Therefore, we also take into account the lens orbital effect, 
%%which is known to produce similar deviations in the light curve to those induced 
%%by the parallax effect \citep{dominik98,jung13}. To account for this effect, we 
%%introduce two linearized parameters $(ds/dt, d\alpha/dt)$, which represent 
%%the relative velocity of the lens components projected onto the plane of the sky. 

We now model the light curve with a set of parameters 
$(t_{0},u_{0},t_{\rm E},s,q,\alpha,\rho,\pi_{{\rm E},N},\pi_{{\rm E},E},ds/dt,d\alpha/dt)$ 
and the color constraint described above. For each of the Close and Wide configurations 
obtained from the ground data sets, we first conduct a grid search for the parameters 
$(\pi_{{\rm E},N}, \pi_{{\rm E},E})$ in order to check the four possibilities of the 
$\pivec_{\rm E}$ measurement. We then rerun the MCMC process with various starting points 
identified in the $(\pi_{{\rm E},N}, \pi_{{\rm E},E})$ space. From this, we find that in both 
configurations, all MCMC chains converged to two points (see the second and third columns in 
Table~\ref{table:one} and~\ref{table:two}). That is, they do not suffer from the degeneracy 
between the pair of $[(+,+), (+,-)]$ or $[(-,-), (-,+)]$, and only suffer from the degeneracy 
between the pair of $[(+,+), (-,-)]$. The latter degeneracy is induced by the mirror symmetry 
of source trajectories relative to the binary axis, i.e., the ``ecliptic degeneracy'' 
\citep{jiang04,skowron11}. Finally, we further investigate the solutions by including the lens orbital effects.

\begin{figure}[th]
%%\epsscale{0.9}
\epsscale{0.9}
\plotone{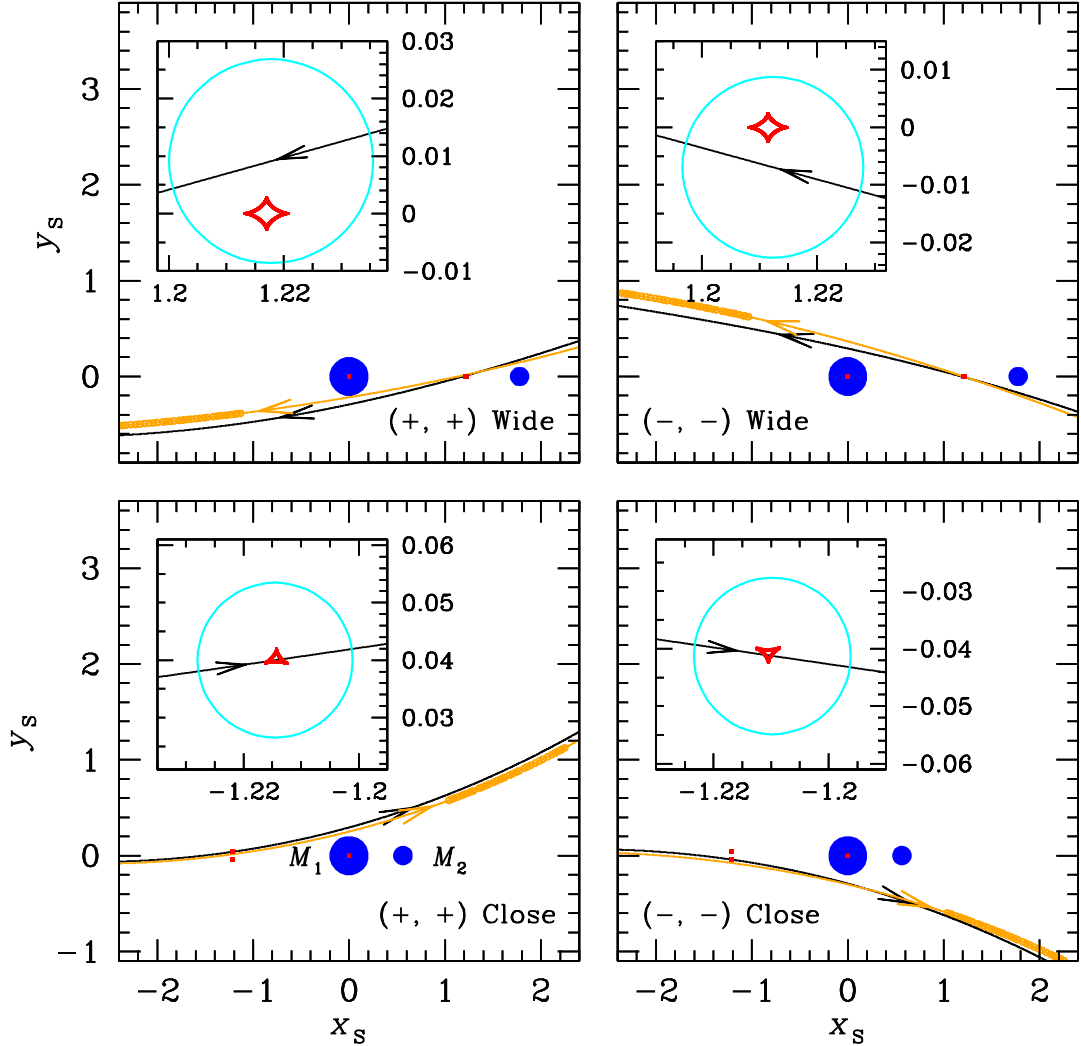}
\caption{\label{fig:three}
Caustic geometries of the four solutions of OGLE-2018-BLG-0596. In each panel, 
the orange curve is the $Spitzer$-viewed source trajectory, while the black curve 
is the Earth-viewed source trajectory. The orange circles are the source positions 
at the times of $Spitzer$ observation. These are not shown to scale in order to 
avoid clutter. The red closed curves are the caustics, and 
the two dark blue dots are the binary lens components. The inset shows the 
enlarged view of the small planetary caustic at the time of the source's caustic 
envelopment. The cyan circle represents the source radius $\rho$ of the best-fit 
solution (cf. Table~\ref{table:one} and~\ref{table:two}).
}
\end{figure}

In Table~\ref{table:one} and~\ref{table:two}, we present the four solutions, 
i.e., $[(+,+), (-,-)]\times({\rm Close},{\rm Wide})$, solutions. The corresponding 
caustic geometries are shown in Figure~\ref{fig:three}. We find that in each solution, 
the source star fully envelops a planetary caustic that is located far from the host. 
In the Close configuration, the anomaly is generated by the envelopment of one of 
the triangular caustics, while in the Wide configuration it is generated by 
the envelopment of the quadrilateral caustic (e.g., \citealt{hwang19}). 
The most important difference between the two sets of solutions is in the mass ratio $q$, 
which is almost 10 times smaller in the Wide solutions.

\subsection{Additional Test for Microlens Parallax}

We find that the $\chi^{2}$ difference of the ground data sets between the standard 
and best-fit $(+,+)_{\rm Close}$ solution is $\Delta\chi^{2} \sim 99$ (see Table~\ref{table:one}). 
This suggests that the microlens parallax is partially constrained by the annual microlens 
parallax effect. To better understand this, we additionally model the ground-based light curve 
with the lens orbital effect and the annual parallax effect (``orbit+AP'').

\begin{figure}[th]
%%\epsscale{0.9}
\epsscale{0.9}
\plotone{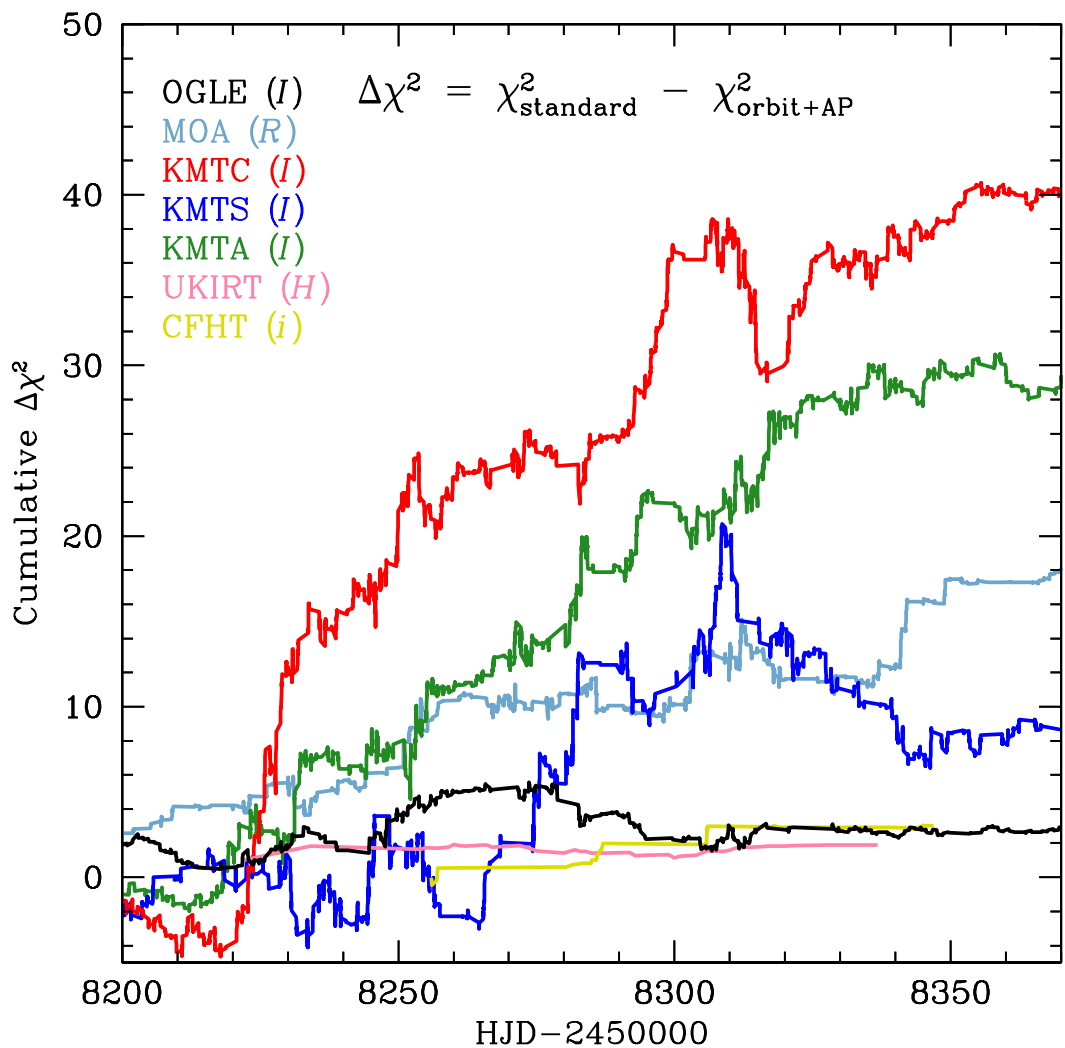}
\caption{\label{fig:four}
Cumulative distribution of $\Delta\chi^{2}$ between the standard 
and the ``orbit+AP'' solution derived from the ground-based observations. 
Note that the diagram is constructed using the best-fit model from 
the $(+,+)_{\rm Close}$ caustic topology.  
}
\end{figure}

However, we find the possibility that the $\chi^{2}$ improvement is caused by systematics of 
the data. From the cumulative distribution of $\Delta\chi^{2} = \chi^{2}_{\rm standard} - \chi^{2}_{\rm orbit+AP}$ 
as a function of time, we find that there is a long-term inconsistent trend between 
KMTNet+MOA and the other data sets (see Figure~\ref{fig:four}). That is, most of 
the improvement comes from the KMTNet+MOA data, while the improvement from the other 
data is very minor. This discrepancy implies that these two data sets may not be 
stable enough to precisely explore the parameter space (e.g., \citealt{han18}). 
From this, one might further conjecture that our $\pivec_{\rm E}$ measurements 
are affected by false-positive effects caused by the systematics. 
We therefore step back and carry out a series of tests to verify our solutions.

First, we fit for the geometric parameters using only the $Spitzer$ data and the 
color constraint. That is, we exclude the ground observations in order to identify 
all possible combinations of $(\pi_{{\rm E},N}, \pi_{{\rm E},E})$ that are consistent 
with the space observations alone. In this modeling, the lensing geometry seen from 
the ground is set by imposing Gaussian constraints on the fitting parameters 
$(t_{0}, u_{0}, t_{\rm E}, f_{{\rm S},{\rm OGLE}}, f_{{\rm B},{\rm OGLE}})$ based on 
the ground-based solution derived in Section 3.1. Next, we include all data sets 
except for KMTNet and MOA, for which we use only partial data sets. While the parallax 
parameters are measured from the overall shape of the light curve, the binary 
parameters are measured from the anomaly. For this event, the overall shape 
is well characterized by the other data sets, but their coverage near the anomaly 
is very poor. To account for the anomaly as well as to remove any spurious parallax 
effects originating from possible systematics, we therefore use only 
the data near the anomaly region (specifically $8235 < {\rm HJD}' < 8250$) 
for KMTNet and MOA. With these modified data sets, we then run full MCMC chains 
incorporating all models obtained from the first step.

\begin{figure}[th]
%%\epsscale{0.9}
\epsscale{1.01}
\plotone{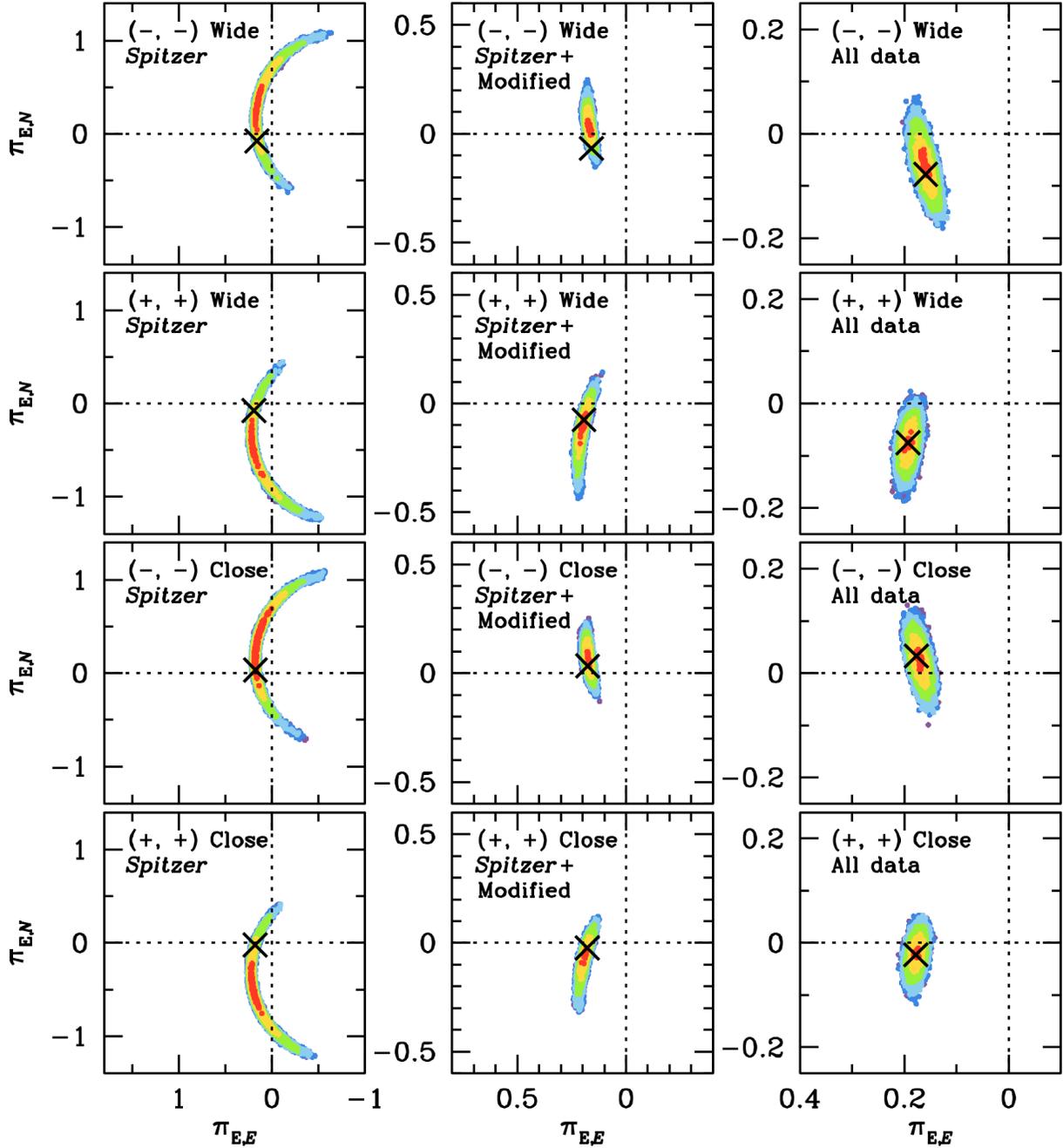}
\caption{\label{fig:five}
$\chi^2$ as a function of microlens parallax for different data sets. 
The left, middle, and right four panels show the derived $\Delta\chi^{2}$ maps 
in the $(\pi_{{\rm E},N}, \pi_{{\rm E},E})$ plane based on the $Spitzer$ data, 
$Spitzer$+modified ground data, and all data sets, respectively. In each panel, 
the black cross marks the location of $\pivec_{\rm E}$ listed in Table~\ref{table:one} and~\ref{table:two}. 
Except that here $n = 1$, the color coding is same as in Figure~\ref{fig:two}. 
Note the different scales in the three columns.
}
\end{figure}

From this test, we find that all MCMC chains converged to two points for both 
the Close and Wide configurations. In addition, the locations of these two points 
in each configuration are nearly identical to those derived from the full 
data sets, indicating that the measured parallaxes are consistent with each other. 
This consistency can be seen in Figure~\ref{fig:five}, where we show the $\Delta\chi^{2}$ 
maps in the $(\pi_{{\rm E},N}, \pi_{{\rm E},E})$ plane obtained from the test. 
We note that the cross mark in each panel represents the location of $\pivec_{\rm E}$ 
listed in Table~\ref{table:one} and~\ref{table:two}, i.e., the four solutions. 
From this figure, one also finds that only the $(+,+)$ and $(-,-)$ models are 
permitted by the modified ground-based data sets. Therefore, we conclude that 
our four solutions are not significantly affected by systematics.

\subsection{Close/Wide Degeneracy}

The $\chi^{2}$ difference between the $(+,+)_{\rm Close}$ and $(+,+)_{\rm Wide}$ 
solution is $\Delta\chi^{2} \sim 17$ \footnote{We note that 
the best-fit solution in each configuration is the $(+,+)_{\rm Close}$ and $(+,+)_{\rm Wide}$ 
solution, and so we consider these two solutions as the representatives.}. 
Mathematically, this implies that the probability of $(+,+)_{\rm Wide}$ solution 
relative to the $(+,+)_{\rm Close}$ solution is lowered by 
$P_{\rm lc} = e^{-\Delta\chi^{2}/2} \sim 2\times10^{-4}$. 
However, this relative fit probability depends on the assumption that all data have 
uncorrelated errors. Unfortunately, such conditions are generally not satisfied for 
crowded field photometry. Hence, it is difficult to entirely reject the 
$(+,+)_{\rm Wide}$ solution from the measured $\Delta\chi^{2}$ alone.

\begin{figure}[th]
%%\epsscale{0.9}
\epsscale{0.9}
\plotone{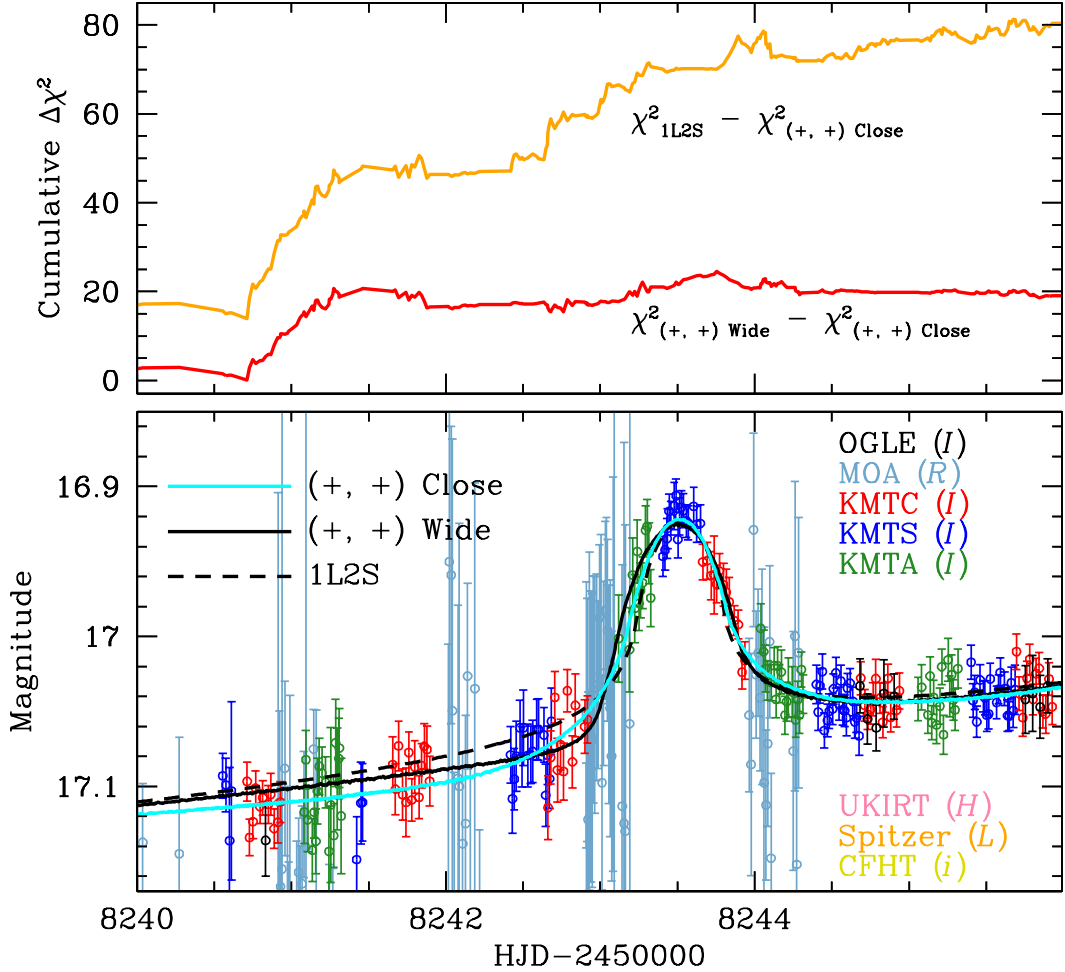}
\caption{\label{fig:six}
Cumulative distributions of $\Delta\chi^{2}$ in the anomaly region. In the upper panel, 
the red and orange curves represent the $\chi^{2}$ differences between the $(+,+)_{\rm Wide}$ 
and $(+,+)_{\rm Close}$ solution and between the 1L2S and $(+,+)_{\rm Close}$ solution, 
respectively. In the lower panel, the cyan, black solid, and black dashed curves 
plotted over the data are the model curves based on the $(+,+)_{\rm Close}$, $(+,+)_{\rm Wide}$, 
and 1L2S solution, respectively.     
}
\end{figure}

Nevertheless, we can better understand the $\chi^{2}$ difference by inspecting 
the cumulative distribution of $\Delta\chi^{2}  = \chi^{2}_{(+,+)_{\rm Wide}} - \chi^{2}_{(+,+)_{\rm Close}}$ 
(see Figure~\ref{fig:six}). From this, we find that most of the difference comes from the 
rising part of the anomaly $({\rm HJD}' \sim 8241)$, where the $(+,+)_{\rm Wide}$ solution 
provides a relatively poor fit to the data. In this region, the $(+,+)_{\rm Wide}$ model 
curve is on average located 0.01 mag above that of the $(+,+)_{\rm Close}$ solution due to 
the difference in the lensing magnification field. For the Close configuration, 
the source passes over the negative planet-host axis during the few days immediately 
prior to the planetary-caustic anomaly (see Figure~\ref{fig:three}). Generically, this axis is 
characterized by a trough (e.g., \citealt{gaudi12}). By contrast, the Wide configuration has 
no such a trough. Moreover, the short-term deviation that favors the Close solution 
cannot be ascribed to the type of long-term systematics discussed above. 
Therefore, we consider that the Wide solutions are disfavored. 
We will further discuss this preference of the data in Section 4.2.

\subsection{Single-lens Binary-source Model}

As discussed by \citet{gaudi98}, short-term anomalies 
can also be produced by a binary source, i.e., 1L2S event. In particular, 
if the binary source (denoted as ``S1'' and ``S2'') has a large flux ratio 
$q_{F} = f_{{\rm S2}}/f_{{\rm S1}}$ and the second source passes very close 
to the lens, the resulting light curve can take a similar form to that of a 
2L1S planetary event. We therefore search for 1L2S solutions based on the 
method of \citet{jung17}. In this search, we simultaneously consider the 
parallax effect, finite-source effect, and orbital motion of the binary source 
(the xallarap effect). We find that the best-fit 1L2S solution is disfavored 
by $\Delta\chi^{2} \sim 87$. To check the result, we also draw the cumulative 
$\Delta\chi^{2}$ distribution of the 1L2S solution relative to the best-fit 2L1S 
solution. As shown in Figure~\ref{fig:six}, we find that most of the $\chi^{2}$ 
difference comes from the short-lived anomaly region $(8240 < {\rm HJD}' < 8245)$, 
in which the 1L2S solution continuously fails to fit the observed light curve. 
Hence, we exclude the 1L2S solution.

\section{Lens Parameters}

The lens total mass $M$ and distance $D_{\rm L}$ can be determined from $\pi_{\rm E}$ 
and $\theta_{\rm E}$ (Equation (3)). These enable us to derive the individual 
masses of the binary lens and their projected separation $a_{\bot} = sD_{\rm L}\theta_{\rm E}$ 
from the measured mass ratio $q$ and separation $s$. In addition, if the source proper 
motion $\muvec_{\rm S}$ is measured, we can derive the lens proper motion 
$\muvec_{\rm L}$ from the relative lens-source proper motion $\muvec_{\rm rel}$ 
(Equation (4)). Then, the lens proper motion can be used to precisely constrain 
the lens physical properties. For the event OGLE-2018-BLG-0596, the proper motion 
of the microlensed source is independently measured from the $Gaia$ observation 
(the $Gaia$ Data Release 2 ID is 4056540540298891520). As will be discussed below, 
this measurement provides us an additional opportunity to investigate the degeneracy 
between our four solutions.

The microlens parallax is measured from the model, while the Einstein radius can be 
measured from $\theta_{\rm E} = \theta_{*}/\rho$. Therefore, we first need to 
determine the angular source radius $\theta_{*}$.

\subsection{Angular Source Radius}

\begin{figure}[th]
%%\epsscale{0.9}
\epsscale{0.9}
\plotone{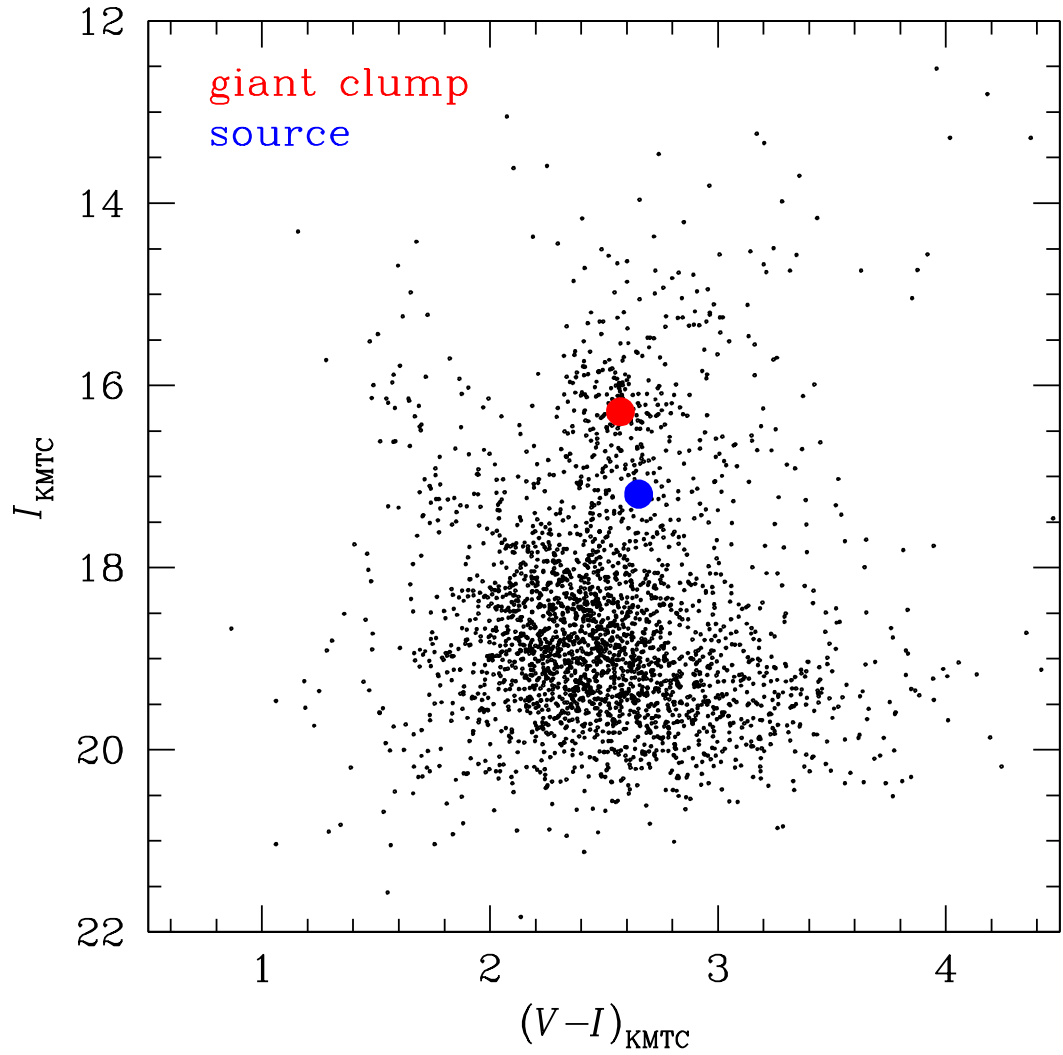}
\caption{\label{fig:seven}
Instrumental Color-magnitude diagram of stars around OGLE-2018-BLG-0596 
(within $120''$) based on the KMTC star catalog. The red and blue dots 
indicate the positions of the giant clump centroid (GC) and the microlensed 
source, respectively.
}
\end{figure}

We evaluate $\theta_{*}$ using the method of \citet{yoo04}. Based on the KMTC star 
catalog constructed by the pyDIA reduction, we first estimate the instrumental source 
color $(V-I)_{\rm S} = 2.65 \pm 0.01$ and magnitude $I_{\rm S} = 17.19 \pm 0.01$ 
from regression and the model, respectively. We next measure the centroid of the 
giant clump (GC) as $(V-I, I)_{\rm GC} = (2.57 \pm 0.02, 16.29 \pm 0.02)$. Figure~\ref{fig:seven} 
displays the locations of the source and GC in the KMTC CMD. We then compare this 
centroid to the calibrated centroid of $(V-I, I)_{0,{\rm GC}} = (1.06 \pm 0.07, 14.40 \pm 0.09)$ 
obtained from \citet{bensby13} and \citet{nataf13}, respectively. This yields an offset 
$\Delta(V-I, I) = (1.51 \pm 0.07, 1.89 \pm 0.09)$. Using this offset, we estimate the 
de-reddened source position as 
\begin{equation}
(V-I, I)_{0,{\rm S}} = (V-I, I)_{\rm S} - \Delta(V-I, I) = (1.14 \pm 0.07, 15.30 \pm 0.09). 
\label{eq10}
\end{equation}
We then apply $(V-I)_{0,{\rm S}}$ to the $VIK$ relation of \citet{bessell88} and derive $(V-K)_{0,{\rm S}} = 2.63\pm0.07$. 
Finally, we estimate $\theta_{*}$ from the $(V-K)_{0,{\rm S}}-\theta_{*}$ relation of \citet{kervella04}, i.e., 
\begin{equation}
\theta_{*} = 4.46 \pm 0.38~\mu{\rm as},
\label{eq11}
\end{equation}
where the error is primarily due to the uncertainty of the GC position and 
color/surface-brightness conversion. The derived source star properties are 
listed in Table \ref{table:three}.

\subsection{Source Proper Motion and Galactic Prior}

The source star of OGLE-2018-BLG-0596 is bright (as derived above), and the blend flux 
associated with the lensing phenomenon is negligible (see Table~\ref{table:one} and~\ref{table:two}). 
In this case, the proper motion measured by $Gaia$ can be attributed to that of the source. 
Then, we can use this measurement to estimate the relative probability of our four solutions 
by comparing the lens projected velocity (from the model) with that expected from the known 
Galactic velocity distribution.

\begin{figure}[th]
%%\epsscale{0.9}
\epsscale{0.9}
\plotone{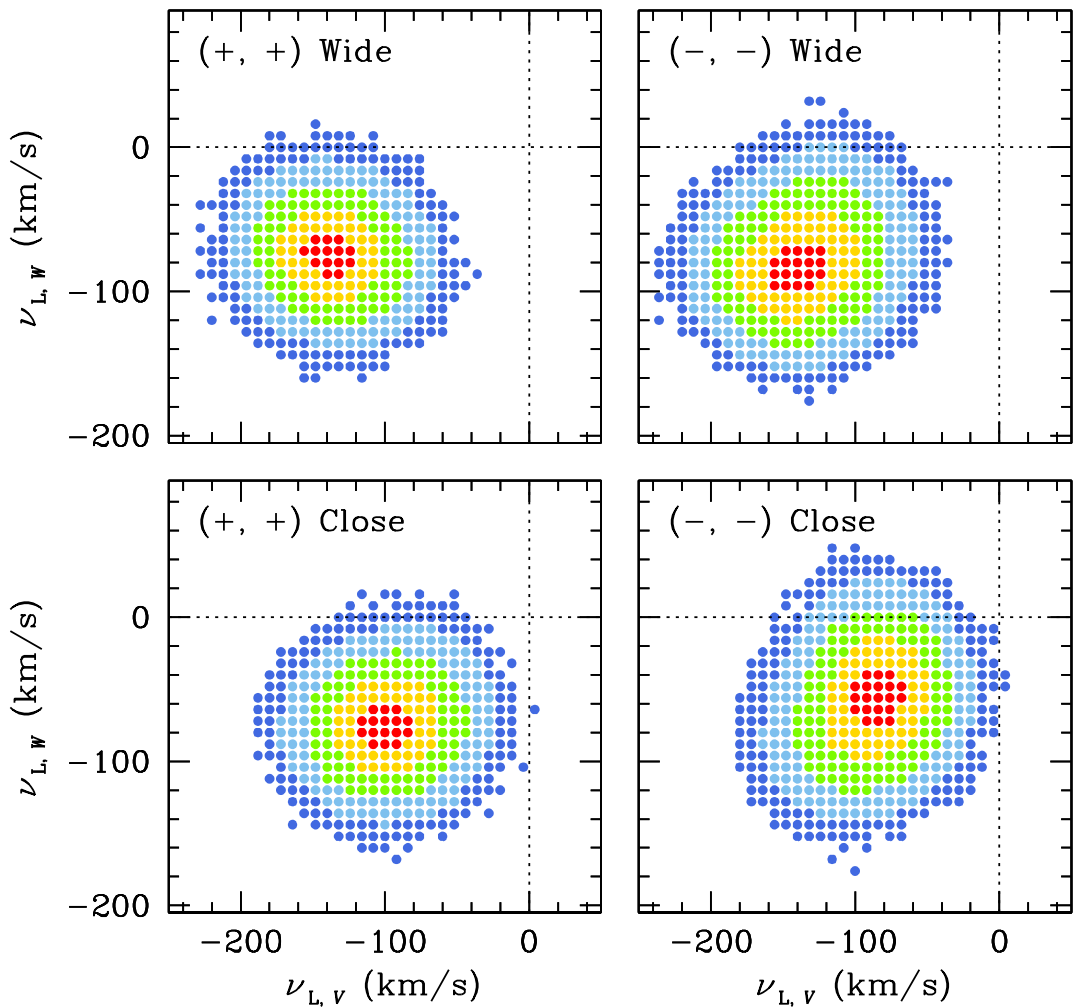}
\caption{\label{fig:eight}
Distributions of lens projected velocities $\nuvec_{\rm L}$ estimated from the four solutions. 
The color notation is same as in Figure~\ref{fig:five}. 
Note that the reference frame is the local standard of rest (LSR). 
}
\end{figure}

For this comparison, we first estimate the lens projected velocity using the four MCMC 
chains summarized in Table~\ref{table:one} and~\ref{table:two}. For each chain, we first 
derive the angular Einstein radius $\theta_{\rm E}$ based on the measured $\theta_{*}$. 
We next estimate the relative lens-source proper motion in the geocentric frame (Equation (3)), 
and transform it to the heliocentric frame, i.e.,    
\begin{equation}
\muvec_{\rm rel, \rm hel} = \muvec_{\rm rel} + \nuvec_{\oplus, \perp}{\pi_{\rm rel} \over {\rm au}}, 
\label{eq12}
\end{equation}
where $\nuvec_{\oplus, \perp} = (\nu_{\oplus, N}, \nu_{\oplus, E}) = (0.47, 28.61)~{\rm km}\,{\rm s}^{-1}$ 
is the projected velocity of Earth at $t_{0}$ \citep{gould04}. The lens proper motion in the heliocentric 
frame is then given by
\begin{equation}
\muvec_{\rm L, \rm hel} = \muvec_{\rm rel, \rm hel} + \muvec_{\rm G}
\label{eq13}
\end{equation}
where $\muvec_{\rm G}(N,E) = (-5.26 \pm 0.55, -4.92 \pm 0.66)~{\rm mas}\,{\rm yr}^{-1}$ 
is the source proper motion measured from $Gaia$ \citep{luri18}. We then estimate the 
lens proper motion $\muvec_{\rm L, {\rm gal}}$ in Galactic coordinates 
with the coordinate transform of \citet{bachelet18}. We finally derive the lens 
projected velocity relative to the local standard of rest (LSR) by 
\begin{equation}
\nuvec_{{\rm L}} = \muvec_{{\rm L}, {\rm gal}}D_{\rm L} + \nuvec_{\odot,{\rm pec}}
\label{eq14}
\end{equation}
where $\nuvec_{\odot,{\rm pec}} = (\nu_{\odot,W}, \nu_{\odot,V}) = (7, 12)~{\rm km}\,{\rm s}^{-1}$ \citep{schonrich10} 
is the peculiar motion of the Sun relative to the LSR\footnote{To estimate $D_{\rm L}$, we generate 
a large number of $D_{\rm S}$ based on a distance distribution drawn from the density profile 
of the Galactic bulge (e.g., \citealt{jung18b}).}. In Figure~\ref{fig:eight}, we show the distributions 
of $\nuvec_{{\rm L}}$ obtained from the four MCMC chains. We note that the $W$ and $V$ axis are defined 
in Cartesian coordinates so that the components point in the direction of the north Galactic pole 
and the Galactic rotation, respectively.

Next, we construct the Galactic velocity distribution in the LSR frame 
based on the model of \citet{robin03}. Because the lens distance is $D_{\rm L} \sim 6~{\rm kpc}$ 
(as derived below), it is expected that the lens is located in the Galactic disk or outer bulge. 
Therefore, we separately consider the bulge, thin disk, and thick disk distributions. We use 
$\bar{\nu}_{{\rm gal}, W} = (\bar{\nu}_{{\rm bulge},W}, \bar{\nu}_{{\rm thin},W}, \bar{\nu}_{{\rm thick},W}) = (0, 0, 0)~{\rm km}\,{\rm s}^{-1}$ and 
$\bar{\sigma}_{{\rm gal},W} = (\bar{\sigma}_{{\rm bulge},W}, \bar{\sigma}_{{\rm thin},W}, \bar{\sigma}_{{\rm thick},W}) = (100, 20, 42)~{\rm km}\,{\rm s}^{-1}$ for the $W$-direction and 
$\bar{\nu}_{{\rm gal},V} = (\bar{\nu}_{{\rm bulge},V}, \bar{\nu}_{{\rm thin},V}, \bar{\nu}_{{\rm thick},V}) = (-220, 0, 0)~{\rm km}\,{\rm s}^{-1}$ and 
$\bar{\sigma}_{{\rm gal},V} = (\bar{\sigma}_{{\rm bulge},V}, \bar{\sigma}_{{\rm thin},V}, \bar{\sigma}_{{\rm thick},V}) = (115, 30, 51)~{\rm km}\,{\rm s}^{-1}$ for the $V$-direction 
with the asymmetric drift of $({\nu}_{{\rm ad},{\rm bulge}}, {\nu}_{{\rm ad},{\rm thin}}, {\nu}_{{\rm ad},{\rm thick}}) = (0, 0, -53)~{\rm km}\,{\rm s}^{-1}$.

In each solution, we separately apply three model distributions to the $k$th chain link 
and derive a probability that the lens has the projected velocity expected from the model distribution, i.e.,  
\begin{equation}
P_{{\rm gal}, k} = \left[{e^{-(\nu_{{\rm L},W} - \nu_{\rm gal,W})^{2}/2\bar{\sigma}_{{\rm gal},W}^{2}}e^{-(\nu_{{\rm L}, V} - \nu_{{\rm gal},V})^{2}/2\bar{\sigma}_{{\rm gal},V}^{2}} \over {\bar{\sigma}_{{\rm gal},W}\bar{\sigma}_{{\rm gal},V}}}\right]\rho_{\rm d, gal}(D_{\rm L})\rho_{\rm d, bulge}(D_{\rm S})
\label{eq15}
\end{equation}
where $\nuvec_{\rm gal} = \bar{\nuvec}_{\rm gal} + \nuvec_{\rm ad}$ 
and $\rho_{\rm d, gal} = (\rho_{\rm d, bulge}, \rho_{\rm d, thin}, \rho_{\rm d, thick})$ 
is the Galactic density profile presented in \citet{jung18b}. 
We then estimate the three probabilities $(P_{\rm bulge}, P_{\rm thin}, P_{\rm thick})$ 
by $P_{\rm gal} = {\Sigma}P_{{\rm gal}, k}$, and find the total probability $P_{\rm tot, gal}$ 
by combining these three probabilities, i.e., $P_{\rm tot, gal} = P_{\rm bulge} + P_{\rm thin} + P_{\rm thick}$. 
Finally, we derive the the net relative probability $P_{\rm net}$ by multiplying the fit probability 
$P_{\rm lc} = e^{-(\chi^{2} - \chi^{2}_{{\rm best}})/2}$ by $P_{\rm tot, gal}$.

The results are listed in Table~\ref{table:four}. 
We find that in both the Close and Wide configurations, the lens system favors the 
bulge populations. This is mainly because the direction of lens projected velocity 
$\nuvec_{\rm L}$ is opposite with respect to the LSR and its magnitude is relatively 
high compared to the rotational velocity $v_{\rm rot} = 220~{\rm km}\,{\rm s}^{-1}$ 
(see Figure~\ref{fig:eight}). From Table~\ref{table:four}, we also find that 
the Galactic-model probabilities $P_{\rm tot, gal}$ are comparable to each other 
and do not lend significant weight to either solution, implying that 
the preference for the Close solutions discussed in Section 3.4 is not significantly 
affected by the Galactic prior. Therefore, we can consider that from 
the balance of evidence $P_{\rm net}$, the Close configuration is strongly favored 
although the Wide configuration cannot be completely ruled out.       

\subsection{Physical Parameters}

For each solution, we now estimate the physical parameters $a_{j}$ 
by imposing the Galactic prior. In the $k$th set of MCMC parameters, 
we evaluate the physical parameters $a_{j,k}$ with the measured $\theta_{*}$ 
and weight them by the probability $P_{{\rm gal}, k}$. We then derive the mean 
and $68\%$ uncertainty range of $a_{j}$ using all weighted $a_{j,k}$. 

The results are listed in Table~\ref{table:five}, which 
includes the lens physical properties $(M_{1}, M_{2}, a_{\perp})$ 
and the event's phase-space coordinates $(D_{\rm L}, D_{\rm S}, \mu_{\rm rel}, \muvec_{\rm rel, hel}, \phi, \muvec_{\rm L, hel}, \nuvec_{\rm L})$. 
Here, $\phi$ is the orientation angle of $\muvec_{\rm rel, hel}$ as measured north through east. 
To investigate the physical validity of these measurements, 
we also show the ratio of the projected kinetic to potential energy \citep{dong09}, 
i.e., $({\rm KE/PE})_{\bot}$. For all four solutions, the low values of $\mu_{\rm rel}$ 
and the large values of $D_{\rm L}$ favor the bulge lenses as predicted 
from the Galactic prior. However, the estimated lens masses for the Close 
and Wide configuration differ from each other due primarily to the difference 
in mass ratios $q$ between the two configurations, but also, to a much smaller 
degree, because of the difference in the normalized source radii $\rho$.

The most favored $(+,+)_{\rm Close}$ solution suggests that the host is 
a mid M-dwarf star with $M_{1} = 0.23\pm0.03~M_{\odot}$, and that the companion 
is a planet with $M_{2} = 13.93\pm1.56~M_{\oplus}$. The projected planet-host 
separation is $a_{\bot} = 0.97\pm0.13~{\rm au}$. Hence, this interpretation indicates 
that the planet is a cold Uranus lying projected outside the snow-line distance, i.e., 
$a_{sl} = 2.7(M_{1}/M_{\odot}) \sim 0.62~{\rm au}$. On the other hand, the $(+,+)_{\rm Wide}$ 
solution corresponds to an Earth-mass planet $(M_{2} = 1.19\pm0.16~M_{\oplus})$ 
orbiting a late M-dwarf $(M_{2} = 0.15\pm0.02~M_{\odot})$. This planet is colder 
because the projected separation $(a_{\bot} = 2.77\pm0.37~{\rm au})$ is 
about $7$ times larger than the snow line.

\section{Discussion} 

We have analyzed the microlensing event OGLE-2018-BLG-0596, which was simultaneously observed from the ground and $Spitzer$. 
The planetary signal in the light curve was densely covered by the data from the KMTNet survey, from which 
the normalized source radius was precisely measured. The $Spitzer$ observations allowed us to measure 
the microlens parallax through the space-based microlens parallax effect. Combined with the source proper 
motion from $Gaia$, these measurements made it possible to precisely constrain the lens physical properties.     

Analysis of the event yields four degenerate solutions originating from two different caustic topologies, 
i.e., $(\pm,\pm)_{\rm Close}$ and $(\pm,\pm)_{\rm Wide}$ solutions. This Close/Wide degeneracy is generated by the 
bright source that fully envelops either the minor-image planetary caustic (Close) or the major-image planetary 
caustic (Wide), i.e., a ``Hollywood'' degeneracy. As pointed out by \citet{hwang19}, the Hollywood degeneracy 
in principle can be resolved because in Close solutions the source passes over the minor-image perturbation 
region, thereby causing a ``dip'' in the light curve near the planetary caustic. In the present case, however, 
the ``dip'' is relatively weak compared to photometric errors. Hence, the degeneracy is only resolved by 
$\Delta\chi^{2} \sim 17$. This $\chi^{2}$ difference is large enough to strongly favor the Close solutions 
but not enough to completely rule out the Wide solutions.

Nevertheless, the degeneracy may be decisively resolved by adaptive optics (AO) follow-up after waiting a time 
for the position of the lens and the microlensed source to separate. This is because the three reasonably competitive 
solutions $[(+,+)_{\rm Close}, (-,-)_{\rm Close}, (+,+)_{\rm Wide}]$ have different heliocentric motion directions 
$\phi = (96\pm11, 81\pm11, 109\pm12)$ deg. For example, if the observed value is $\phi_{\rm AO} = 80$ deg, this will 
strongly favor the Close solutions because it is inconsistent with that of the Wide solution. If the value is 
$\phi_{\rm AO} = 90$ deg, then this would marginally disfavor the Wide solution. However, given the strong $\chi^{2}$ 
preference for the Close solutions, this would still clearly resolve the degeneracy. 

The best-fit $(+,+)_{\rm Close}$ solution has the planet-host mass ratio of $q =1.8\times10^{-4}$, 
which is just larger than the peak in the mass ratio distribution of \citet{jung19a} that suggests 
a pile-up of Neptune-mass planets. However, even though the mass ratio is near the middle of a 
predicted ``gap'' between Neptune- and Jupiter-mass planets for Solar-mass hosts, the derived physical 
solution has an ice-giant planet with $M_{\rm p} = 13.93\pm1.56~M_{\oplus}$, very similar to Uranus 
in our solar system. This is because the lens host is a mid M-dwarf whose mass is much smaller than 
the Solar mass, i.e., $M_{\rm h} = 0.23\pm0.03~M_{\odot}$. This implies that one needs to be cautious 
about interpreting the continuous mass ratio distribution (e.g., \citealt{jung19a}) as indicating a 
continuous planet mass distribution. That is, we need precise host masses for the many microlensing planets 
without $\pivec_{\rm E}$ measurements in order to correctly understand the planet distribution beyond 
the snow line. Such host-mass measurements will be possible for the majority of microlensing planets 
detected to date at first AO light on next-generation (30 m) telescopes.

\acknowledgments
This research has made use of the KMTNet system operated by the Korea 
Astronomy and Space Science Institute (KASI) and the data were obtained at 
three host sites of CTIO in Chile, SAAO in South Africa, and SSO in Australia. 
Work by AG was supported by AST-1516842 from the US NSF. AG was 
supported by JPL grant 1500811. AG received support from the European Research Council 
under the European Unions Seventh Framework Programme (FP 7) ERC Grant Agreement n. [321035]. 
Work by C. Han was supported by grant 2017R1A4A1015178 of the National Research Foundation of Korea. 
Work by MTP was partially supported by NASA grants NNX16AC62G and NNG16PJ32C. 
The OGLE has received funding from the National Science Centre, Poland, grant MAESTRO 
2014/14/A/ST9/00121 to A.U. The MOA project is supported by JSPS KAKENHI grants 
No. JSPS24253004, JSPS26247023, JSPS23340064, JSPS15H00781, JP16H06287,and JP17H02871. 
UKIRT is currently owned by the University of Hawaii (UH) and operated by the 
UH Institute for Astronomy; operations are enabled through the cooperation of the East 
Asian Observatory. The collection of the 2018 data reported here was supported by NASA grant 
NNG16PJ32C and JPL proposal \#18-NUP2018-0016. This research uses data obtained through 
the Telescope Access Program (TAP), which has been funded by the National Astronomical 
Observatories, Chinese Academy of Sciences, and the Special Fund for Astronomy from 
the Ministry of Finance. The work by WZ, TW and SM is partly supported by the National 
Science Foundation of China (Grant No. 11821303 and 11761131004 to SM).

% Table 1 ------------------------------------------------
%\begin{deluxetable*}{lrrrrr}{ht}
%%\begin{landscape}
\begin{deluxetable}{lrrrrrr}
%%\tabletypesize{\scriptsize}
\tabletypesize{\footnotesize}
\tablecaption{Lensing Parameters for Close Solutions\label{table:one}}
\tablewidth{0pt}
\tablehead{
\multicolumn{1}{l}{Parameters} &
\multicolumn{1}{c}{Standard} &
\multicolumn{2}{c}{Parallax} &
\multicolumn{2}{c}{Orbit+Parallax} \\
\multicolumn{1}{c}{} & 
\multicolumn{1}{c}{} & 
\multicolumn{1}{c}{$(+, +)$} & 
\multicolumn{1}{c}{$(-, -)$} &
\multicolumn{1}{c}{$(+, +)$} & 
\multicolumn{1}{c}{$(-, -)$} 
}
\startdata
$\chi^2_{\rm tot}$/dof     &   25959.6/26562     &   25901.6/26596      &   25906.6/26596       &    25895.4/26594      &   25897.3/26594           \\
$t_0$ (${\rm HJD'}$)       & 8277.16$\pm$ 0.010  & 8277.14$\pm$ 0.010   & 8277.14$\pm$ 0.010    & 8277.14$\pm$ 0.015    & 8277.14 $\pm$ 0.015       \\
$u_0$                      &  0.285 $\pm$ 0.002  &  0.285 $\pm$ 0.004   & -0.284 $\pm$ 0.004    &  0.284 $\pm$ 0.006    &  -0.282 $\pm$ 0.006        \\
$t_{\rm E}$ (days)         & 28.881 $\pm$ 0.109  & 28.924 $\pm$ 0.110   & 29.038 $\pm$ 0.109    & 29.003 $\pm$ 0.110    &  29.170 $\pm$ 0.116        \\
$s$                        &  0.566 $\pm$ 0.003  &  0.564 $\pm$ 0.005   &  0.565 $\pm$ 0.005    &  0.512 $\pm$ 0.017    &   0.499 $\pm$ 0.018        \\
$q$ ($10^{-4}$)            &  1.203 $\pm$ 0.089  &  1.327 $\pm$ 0.105   &  1.313 $\pm$ 0.106    &  1.827 $\pm$ 0.132    &   1.879 $\pm$ 0.133        \\
$\alpha$ (rad)             &  6.072 $\pm$ 0.009  &  6.076 $\pm$ 0.011   & -6.071 $\pm$ 0.011    &  6.022 $\pm$ 0.011    &  -6.025 $\pm$ 0.014        \\
$\rho_\ast$ ($10^{-2}$)    &  1.120 $\pm$ 0.042  &  1.139 $\pm$ 0.043   &  1.137 $\pm$ 0.042    &  1.347 $\pm$ 0.049    &   1.362 $\pm$ 0.050        \\
$\pi_{{\rm E}, N}$         & --                  & -0.041 $\pm$ 0.023   &  0.043 $\pm$ 0.022    & -0.023 $\pm$ 0.022    &   0.033 $\pm$ 0.026         \\
$\pi_{{\rm E}, E}$         & --                  &  0.177 $\pm$ 0.010   &  0.179 $\pm$ 0.010    &  0.178 $\pm$ 0.010    &   0.177 $\pm$ 0.010         \\
$ds/dt$ (yr$^{-1}$)        & --                  & --                   & --                    & -0.580 $\pm$ 0.045    &  -0.734 $\pm$ 0.031         \\
$d\alpha/dt$ (yr$^{-1}$)   & --                  & --                   & --                    & -0.637 $\pm$ 0.101    &   0.566 $\pm$ 0.135        \\    
$f_{\rm S}$                &  1.956 $\pm$ 0.002  &  1.955 $\pm$ 0.002   &  1.945 $\pm$ 0.002    &  1.945 $\pm$ 0.002    &   1.930 $\pm$ 0.002        \\  
$f_{\rm B}$                & -0.044 $\pm$ 0.003  & -0.043 $\pm$ 0.003   & -0.032 $\pm$ 0.003    & -0.032 $\pm$ 0.003    &  -0.017 $\pm$ 0.003       \\   
$\chi^2_{\rm ground}$      &  25959.6            &  25867.4             &  25871.6              &    25860.9            &   25861.8                 \\
$\chi^2_{Spitzer}$         & --                  &  34.1                &  34.9                 &    34.5               &   35.1                   \\
$\chi^2_{\rm penalty}$     & --                  &  0.11                &  0.14                 &    0.0079             &   0.43                 
\enddata 
\vspace{0.05cm}
%\tablecomments{
%${\rm HJD}'= {\rm HJD}-2450000$
%}
%\end{deluxetable*}
\end{deluxetable}
%%\end{landscape}
% -----------------------------------------------------

% Table 2 ------------------------------------------------
%\begin{deluxetable*}{lrrrrr}{ht}
%%\begin{landscape}
\begin{deluxetable}{lrrrrrr}
%%\tabletypesize{\scriptsize}
\tabletypesize{\footnotesize}
\tablecaption{Lensing Parameters for Wide Solutions\label{table:two}}
\tablewidth{0pt}
\tablehead{
\multicolumn{1}{l}{Parameters} &
\multicolumn{1}{c}{Standard} &
\multicolumn{2}{c}{Parallax} &
\multicolumn{2}{c}{Orbit+Parallax} \\
\multicolumn{1}{c}{} & 
\multicolumn{1}{c}{} & 
\multicolumn{1}{c}{$(+, +)$} & 
\multicolumn{1}{c}{$(-, -)$} &
\multicolumn{1}{c}{$(+, +)$} & 
\multicolumn{1}{c}{$(-, -)$} 
}
\startdata
$\chi^2_{\rm tot}$/dof     &    26009.5/26562    &    25915.7/26596    &    25926.1/26596     &    25912.2/26594      &     25924.0/26594     \\
$t_0$ (${\rm HJD'}$)       & 8277.16$\pm$ 0.010  & 8277.13$\pm$ 0.011  & 8277.14$\pm$ 0.011   & 8277.13$\pm$ 0.015    & 8277.13 $\pm$ 0.014   \\
$u_0$                      &  0.286 $\pm$ 0.002  &  0.284 $\pm$ 0.004  & -0.286 $\pm$ 0.004   &  0.285 $\pm$ 0.006    &  -0.285 $\pm$ 0.005    \\
$t_{\rm E}$ (days)         & 28.833 $\pm$ 0.107  & 28.974 $\pm$ 0.110  & 28.870 $\pm$ 0.105   & 28.889 $\pm$ 0.107    &  28.894 $\pm$ 0.105    \\
$s$                        &  1.769 $\pm$ 0.004  &  1.773 $\pm$ 0.006  &  1.775 $\pm$ 0.006   &  1.811 $\pm$ 0.015    &   1.788 $\pm$ 0.016    \\
$q$ ($10^{-4}$)            &  0.160 $\pm$ 0.015  &  0.187 $\pm$ 0.015  &  0.181 $\pm$ 0.014   &  0.244 $\pm$ 0.027    &   0.188 $\pm$ 0.019    \\
$\alpha$ (rad)             &  2.897 $\pm$ 0.011  &  2.899 $\pm$ 0.018  & -2.901 $\pm$ 0.015   &  2.929 $\pm$ 0.018    &  -2.921 $\pm$ 0.016    \\
$\rho_\ast$ ($10^{-2}$)    &  1.495 $\pm$ 0.066  &  1.587 $\pm$ 0.074  &  1.591 $\pm$ 0.058   &  1.772 $\pm$ 0.078    &   1.572 $\pm$ 0.059    \\
$\pi_{{\rm E}, N}$         & --                  & -0.061 $\pm$ 0.025  & -0.071 $\pm$ 0.024   & -0.075 $\pm$ 0.026    &  -0.078 $\pm$ 0.032     \\
$\pi_{{\rm E}, E}$         & --                  &  0.191 $\pm$ 0.010  &  0.157 $\pm$ 0.010   &  0.193 $\pm$ 0.011    &   0.159 $\pm$ 0.011     \\
$ds/dt$ (yr$^{-1}$)        & --                  & --                  & --                   &  0.371 $\pm$ 0.048    &   0.151 $\pm$ 0.036     \\
$d\alpha/dt$ (yr$^{-1}$)   & --                  & --                  & --                   &  0.355 $\pm$ 0.089    &  -0.225 $\pm$ 0.091    \\    
$f_{\rm S}$                &  1.961 $\pm$ 0.002  &  1.949 $\pm$ 0.002  &  1.961 $\pm$ 0.002   &  1.959 $\pm$ 0.002    &   1.958 $\pm$ 0.002    \\  
$f_{\rm B}$                & -0.048 $\pm$ 0.003  & -0.035 $\pm$ 0.003  & -0.047 $\pm$ 0.003   & -0.045 $\pm$ 0.003    &  -0.045 $\pm$ 0.003   \\   
$\chi^2_{\rm ground}$      &  26009.5            &  25881.5            &  25890.3             &    25878.3            &     25887.5           \\
$\chi^2_{Spitzer}$         & --                  &  34.1               &  35.6                &    33.9               &     35.6             \\
$\chi^2_{\rm penalty}$     & --                  &  0.12               &  0.16                &    0.0019             &     0.94           
\enddata 
\vspace{0.05cm}
%\tablecomments{
%${\rm HJD}'= {\rm HJD}-2450000$
%}
%\end{deluxetable*}
\end{deluxetable}
%%\end{landscape}
% -----------------------------------------------------

% Table 3 ------------------------------------------------
\begin{deluxetable}{lrr}
\tablecaption{Derived Properties for Source Star\label{table:three}}
\tablewidth{0pt}
\tablehead{
\colhead{Parameter}   &
\colhead{Units}       &
\colhead{Value}
}
\startdata
$I_{\rm S}$           &   [mag]       &   17.19 $\pm$ 0.01   \\
$(V-I)_{\rm S}$       &               &   2.65 $\pm$ 0.01    \\
$I_{0,{\rm S}}$       &   [mag]       &   15.30 $\pm$ 0.09   \\
$(V-I)_{0,{\rm S}}$    &               &   1.14 $\pm$ 0.07    \\
$(V-K)_{0,{\rm S}}$    &               &   2.63 $\pm$ 0.07    \\
$\theta_{*}$          &   [$\mu$as]   &   4.46 $\pm$ 0.38
\enddata
\tablenotetext{a}{See \S4.1 for details.}
\end{deluxetable}

% Table 4 ------------------------------------------------
%\begin{deluxetable*}{lrrrrr}{ht}
\begin{deluxetable}{lllll}
\tablecaption{Relative Probabilities\label{table:four}}
\tablewidth{0pt}
\tablehead{
\multicolumn{1}{l}{Parameters} &
\multicolumn{2}{c}{Close} &
\multicolumn{2}{c}{Wide} \\
\multicolumn{1}{c}{} & 
\multicolumn{1}{c}{$(+, +)$} & 
\multicolumn{1}{c}{$(-, -)$} &
\multicolumn{1}{c}{$(+, +)$} & 
\multicolumn{1}{c}{$(-, -)$} 
}
\startdata
$P_{\rm lc}$               &  1.0                     &   0.37       &   $2.04\times10^{-4}$        &   $5.04\times10^{-7}$    \\
$P_{\rm thin}$             &  0.24                    &   8.52       &   $1.21\times10^{-2}$        &   $4.08\times10^{-2}$    \\
$P_{\rm thick}$            &  14.11                   &   40.60      &   8.14                       &   7.39                   \\
$P_{\rm bulge}$            &  78.46                   &   101.19     &   160.04                     &   185.63                 \\
$P_{\rm tot, gal}$         &  92.81                   &   150.31     &   168.19                     &   193.06                 \\      
$P_{\rm net}$              &  92.81                   &   55.61      &   $3.43\times10^{-2}$        &   $9.73\times10^{-5}$ 
\enddata 
\vspace{0.05cm}
%%\tablecomments{
%%${\rm HJD}'= {\rm HJD}-2450000$
%%}
%\end{deluxetable*}
\end{deluxetable}
% -----------------------------------------------------

% Table 5 ------------------------------------------------
%\begin{deluxetable*}{lrrrrr}{ht}
\begin{deluxetable}{lrrrr}
\tablecaption{Physical Parameters\label{table:five}}
\tablewidth{0pt}
\tablehead{
\multicolumn{1}{l}{Parameters} &
%%\multicolumn{2}{c}{Close} &
%%\multicolumn{2}{c}{Wide} \\
%%\multicolumn{1}{c}{} & 
%%\multicolumn{1}{c}{$(+, +)$} & 
%%\multicolumn{1}{c}{$(-, -)$} &
%%\multicolumn{1}{c}{$(+, +)$} & 
%%\multicolumn{1}{c}{$(-, -)$} 
\multicolumn{1}{c}{$(+,+)_{\rm Close}$} &
\multicolumn{1}{c}{$(-,-)_{\rm Close}$} & 
\multicolumn{1}{c}{$(+,+)_{\rm Wide}$} &
\multicolumn{1}{c}{$(-,-)_{\rm Wide}$} 
}
\startdata
$\theta_{\rm E}$ (mas)                                     &   $0.336\pm0.034$  &   $0.324\pm0.033$  &   $0.256\pm0.027$   &   $0.276\pm0.028$  \\ 
$M_{1}$ $(M_{\odot})$                                      &   $0.231\pm0.028$  &   $0.229\pm0.031$  &   $0.154\pm0.019$   &   $0.196\pm0.021$  \\ 
$M_{2}$ $(M_{\oplus})$                                     &   $13.93\pm1.56$   &   $14.69\pm1.72$   &    $1.19\pm0.16$    &    $1.30\pm0.16$   \\ 
$a_{\bot}$ (au)                                            &    $0.97\pm0.13$   &    $0.92\pm0.14$   &    $2.77\pm0.37$    &    $3.02\pm0.39$   \\
$D_{\rm L}$ (kpc)                                          &    $5.65\pm0.75$   &    $5.65\pm0.80$   &    $5.95\pm0.77$    &    $6.11\pm0.79$   \\
$D_{\rm S}$ (kpc)                                          &    $8.58\pm1.42$   &    $8.38\pm1.47$   &    $8.69\pm1.35$    &    $8.67\pm1.33$   \\
$\mu_{{\rm rel}}$ $({\rm mas}\,{\rm yr}^{-1})$             &    $4.22\pm0.42$   &    $4.07\pm0.41$   &    $3.24\pm0.33$    &    $3.49\pm0.35$   \\ 
$\mu_{{\rm rel},{\rm hel},N}$ $({\rm mas}\,{\rm yr}^{-1})$ &   $-0.44\pm0.50$   &    $0.64\pm0.57$   &   $-1.13\pm0.36$    &   $-1.14\pm0.65$   \\ 
$\mu_{{\rm rel},{\rm hel},E}$ $({\rm mas}\,{\rm yr}^{-1})$ &    $4.53\pm0.46$   &    $3.98\pm0.43$   &    $3.33\pm0.36$    &    $3.24\pm0.33$   \\ 
$\phi$ (deg) (E of N)                                      &    $95.5\pm10.9$   &    $81.3\pm10.9$   &   $108.7\pm11.8$    &   $107.4\pm14.3$   \\
$\mu_{{\rm L},{\rm hel},N}$ $({\rm mas}\,{\rm yr}^{-1})$   &   $-5.67\pm0.79$   &   $-4.57\pm0.80$   &   $-6.37\pm0.76$    &   $-6.37\pm0.94$   \\    
$\mu_{{\rm L},{\rm hel},E}$ $({\rm mas}\,{\rm yr}^{-1})$   &   $-0.45\pm0.41$   &   $-0.66\pm0.44$   &   $-1.63\pm0.44$    &   $-1.44\pm0.41$   \\  
$\nu_{{\rm L},W}$ $({\rm km}\,{\rm s}^{-1})$               &   $-98.8\pm18.7$   &   $-73.6\pm22.2$   &   $-97.1\pm16.7$    &  $-103.5\pm20.4$   \\
$\nu_{{\rm L},V}$ $({\rm km}\,{\rm s}^{-1})$               &   $-97.4\pm20.1$   &   $-82.5\pm18.6$   &  $-141.3\pm22.8$    &  $-141.1\pm24.7$   \\
$({\rm KE/PE})_{\bot}$ ($10^{-2}$)                         &    $8.67\pm2.86$   &   $10.06\pm3.22$   &  $50.78\pm21.92$    &   $11.91\pm8.19$  

\enddata 
\vspace{0.05cm}
%%\tablecomments{
%%${\rm HJD}'= {\rm HJD}-2450000$
%%}
%\end{deluxetable*}
\end{deluxetable}
% -----------------------------------------------------

\end{document}